\documentclass[preprint,superscriptaddress]{revtex4}

\usepackage{graphicx}
\usepackage{dcolumn}
\usepackage{amsmath}
\usepackage{epstopdf}
\usepackage[colorlinks=true, citecolor=blue, urlcolor = blue, linkcolor= blue, bookmarks=true]{hyperref}
\begin{document}


\title{Shadows of rotating five-dimensional charged EMCS black holes}

\author{Muhammed Amir}
\email{amirctp12@gmail.com}
\affiliation{Centre for Theoretical Physics,
Jamia Millia Islamia,  New Delhi 110025, India}

\author{Balendra Pratap Singh}
\email{balendra29@gmail.com}
\affiliation{Centre for Theoretical Physics,
Jamia Millia Islamia,  New Delhi 110025, India}

\author{Sushant G. Ghosh}
\email{sghosh2@jmi.ac.in}
\affiliation{Centre for Theoretical Physics,
 Jamia Millia Islamia,  New Delhi 110025, India}
\affiliation{Multidisciplinary Centre for Advanced Research and Studies (MCARS),
 Jamia Millia Islamia,  New Delhi 110025, India}
\affiliation{Astrophysics and Cosmology Research Unit,
 School of Mathematics, Statistics and Computer Science,
 University of KwaZulu-Natal, Private Bag X54001,
 Durban 4000, South Africa}

\begin{abstract}
Higher dimensional theories admit astrophysical objects like supermassive black holes, which 
are rather different from standard ones, and their gravitational lensing features deviate from  
general relativity. It is well known that a black hole shadow is a dark region due to the 
falling geodesics of photons into the black hole and, if detected, a black hole shadow could 
be used to determine which theory of gravity is consistent with observations. Measurements of 
the shadow sizes around the black holes can help to evaluate various parameters of the black 
hole metric. We study the shapes of the shadow cast by the rotating five-dimensional charged 
Einstein-Maxwell-Chern-Simons (EMCS) black holes, which is characterized by the four 
parameters, i.e., mass, two spins, and charge, in which the spin parameters are set equal. We
integrate the null geodesic equations and derive an analytical formula for the shadow of the
five-dimensional EMCS black hole, in turn, to show that size of  black hole shadow is affected 
due to charge as well as spin. The shadow is a dark zone covered by a deformed circle, and the 
size of the shadow decreases with an increase in the charge $q$ when compared with the five-
dimensional Myers-Perry black hole. Interestingly, the distortion increases with charge $q$. 
The effect of these parameters on the shape and size of the naked singularity shadow of 
five-dimensional EMCS black hole is also discussed.
\end{abstract}

\maketitle

\section{Introduction}
Black holes are intriguing astrophysical objects and perhaps the most fascinating objects in 
Universe and it is hard to find any other object or topic that attracts more attention. 
However, it is still not clear whether black holes can be observed. Observation of the shadow 
of black hole candidate Sagittarius A$^*$ or Sgr A$^*$ is one of the most important goals of 
the Very Large Baseline Interferometry (VLBI) technique. It should able to image the  black 
hole with resolution at the level of the event horizon. At galactic center the black hole 
candidate Sgr A$^*$, due to the gravitational lensing effect casts a shadow; the shape and 
size of this shadow can be calculated. There is a widespread belief that evidence for the 
existence of black holes will come from the direct observation of its shadow. The shape of a 
shadow could be used to study extreme gravity near the event horizon and also to know whether 
the general relativity is consistent with the observations. Observation of a black hole shadow 
may allow us to determine the mass and spin of a rotating black hole \cite{Vries, Takahashi04, Takahashi041,Bambi:2008jg,Bambi:2010hf,Goddi:2016jrs}. As black holes are basically non 
emitting objects, it is of interest the study null geodesics around them where photons coming 
from other sources move to lead to a shadow. To a distant observer, the event horizons cast 
shadows due to the bending of light by a black hole  \cite{Falcke:1999pj}. A  first step 
towards the study of a black hole shadow was done by Bardeen \cite{JMB}, who calculated the 
shape of a dark area of a Kerr black hole, i.e., its shadow over a bright background. Although 
the shadow of Schwarzschild black hole is a perfect circle \cite{Synge:1963,Luminet:1979},  
the Kerr black hole  does not have a circular shadow image; it has an elongated shape in the 
direction of rotation \cite{Chandrasekhar}. The pictures of individual spherical light-like 
geodesics in the Kerr spacetime can be found in \cite{Teo}, and one can find a discussion and 
the picture of the photon region in the background of Kerr spacetime in Ref.~\cite{Perlick}.

Thus, the shadow deviation from the circle can determine the spin parameter of black holes. 
The study of black holes has been extended for other black holes, such as Kerr black holes 
\cite{JMB}, Kerr-Newman black holes\cite{Takahashi:2005hy}, regular black holes 
\cite{Li:2013jra,Abdujabbarov:2016hnw,Amir:2016cen}, 
multi-black holes \cite{Yumoto:2012kz}, black holes in extended Chern-Simons modified gravity 
\cite{Amarilla:2010zq} and Randall-Sundrum braneworld \cite{Amarilla:2011fxx} case. The 
shadows of black holes with nontrivial NUT charge were obtained in \cite{Grenzebach:2014fha}, 
while the Kerr-Taub-NUT black holes were discussed in \cite{Abdujabbarov:2012bnn}. The 
apparent shape of the Kerr-Sen black holes is studied in \cite{Hioki:2008zw}, and rotating 
braneworld black holes were investigated in \cite{Amarilla:2011fxx,Schee:2008kz}. Further, the 
effect of the spin parameter on the shape of the shadow was extended to the Kaluza-Klein 
rotating dilaton black hole \cite{Amarilla:2013sj}, the rotating Horava-Lifshitz black hole 
\cite{Atamurotov:2013dpa}, the rotating non-Kerr black hole \cite{Atamurotov:2013sca} and the 
Einstein-Maxwell-dilaton-axion black hole \cite{Wei:2013kza}. There are  different approaches 
to calculating the shadow of the black holes, e.g., a coordinate-independent characterization 
\cite{Abdujabbarov:2015xqa} and general relativistic ray-tracing \cite{Younsi:2016azx}.

Recent years witnessed black hole solutions in more than four spacetime dimensions, especially 
in five-dimensions as the subject of intensive research motivated by ideas in the braneworld, 
string theory and gauge/gravity duality. Several interesting and surprising results have been 
found in \cite{Horowitz:2011c}. The models with large, extra dimensions have been proposed to 
deal with several issues arising in modern particle phenomenology \cite{Horowitz:2011c,ArkaniHamed:1998rs,Antoniadis:1998ig,Randall:1999ee}. The rotating black holes have many 
applications and display interesting structures, but they are also very difficult to find in 
higher dimensions and the bestiary for solutions is much wider and less understood 
\cite{Emparan:2008eg,Adamo:2014baa}. The uniqueness theorems do not hold in higher dimensions 
due to the fact that there are more degrees of freedom. The black-ring solution in five 
dimensions shows that higher-dimensional spacetime can admit nontrivial topologies 
\cite{Reall:2002bh}. The Myers-Perry black hole solution \cite{Myers:1986un} is a 
higher-dimensional generalization of the  Kerr black hole solution. However, the Kerr-Newman 
black hole solution in higher-dimensions has not yet been discovered. Nevertheless, there is a 
related solution of the Einstein-Maxwell-Chern-Simons (EMCS) theory in the five-dimensional 
minimal gauged supergravity \cite{Chong:2005hr,Gauntlett:2002}. Remarkably, the exact five-
dimensional solutions for rotating charged black holes are known in the Einstein-Maxwell 
theory with a Chern-Simons term. The extremal limits of the five-dimensional rotating charged 
black hole solutions  are of special interest, since they encompass a two parameter family of 
stationary supersymmetric black holes. The shadow of five-dimensional rotating black hole 
\cite{Papnoi:2014} suggests that the shadow is slightly smaller and less deformed than for its 
four-dimensional Kerr black hole counterpart. Recently, the shadow of higher-dimensional 
Schwarschild-Tangherlini black hole was discussed in \citep{Singh:2017vfr} and the results 
show that the size of the shadows decreases in higher-dimensions.

The aim of this paper is to investigate the shadow of a five-dimensional EMCS minimal gauged 
supergravity black hole (henceforth five-dimensional EMCS black hole) and compare the results 
with the Kerr black hole/five-dimensional Myers-Perry black hole. We have discussed in detail 
the structure of the horizons and the shadows of EMCS black holes in five dimensions, focusing 
on solutions with equal magnitude angular momenta, to enhance the symmetry of the solutions, 
making the analytical analysis much more tractable, while at the same time revealing already 
numerous intriguing features of the black hole shadows.

The paper is organized as follows. In Sect.~\ref{EMCS_BH}, we review the five-dimensional 
Myers-Perry black hole solution and also visualize the ergosphere for various values of the 
charge $q$. In Sect.~\ref{pm}, we present the particle motion around the five-dimensional EMCS 
black hole to discuss the black hole shadow. Two observables are introduced to discuss the 
apparent shape of the black hole shadow in Sect.~\ref{5DShd}. The naked singularity shadow of 
five-dimensional EMCS spacetime is subject of Sect.~\ref{naksing}. We discuss the energy 
emission rate of five-dimensional EMCS black hole in Sect.~\ref{emission} and finally  we 
conclude the main results in Sect.~\ref{conclusion}.

We have used units that fix the speed of light and the gravitational constant via 
$8\pi G=c=1$.

\section{Rotating five-dimensional EMCS black holes}
\label{EMCS_BH}
It is well known that the stationary black holes in five-dimensional EMCS theory possess 
surprising properties when considering the Chern-Simons coefficient as a parameter. We 
briefly review the five-dimensional asymptotically flat EMCS black hole solutions. 
The Lagrangian for the bosonic sector of the minimal five-dimensional supergravity reads 
\cite{Cremmer:1981,Reimers:2016czc}: 
\begin{equation}\label{lagrange}
{\cal L}= \frac{1}{16\pi } \left[\sqrt{-g}(R -F^2) -
\frac{2}{3\sqrt{3}}\epsilon^{\mu \nu \rho \sigma \tau}A_\mu F_{\nu \rho}F_{\sigma \tau}\right],
\end{equation}
where $R$ is the curvature scalar, $ F_{\mu \nu} = \partial_\mu A_\nu -\partial_\nu A_\mu $ 
with $A_{\mu}$ is the gauge potential, and $\epsilon^{\mu \nu \lambda \rho \sigma}$ is the 
five-dimensional Levi-Civita tensor. The Lagrangian (\ref{lagrange}) has an additional 
Chern-Simons term different from  the usual Einstein-Maxwell term. The corresponding equations 
of motion reads \cite{Reimers:2016czc,Maldacena:1997re}
\begin{eqnarray}
R_{\mu \nu}-\frac{1}{2}g_{\mu \nu}R = 2\left(F_{\mu \alpha}F_{\nu \alpha}
-\frac{1}{4}g_{\mu \nu}F_{\rho \sigma}F^{\rho \sigma}\right), \nonumber\\
\nabla_{\mu} \left(F^{\mu \nu}
+\frac{1}{\sqrt{3} \sqrt{-g}}\epsilon^{\mu \nu \lambda \rho \sigma}
A_{\lambda} F_{\rho \sigma}\right) = 0.
\end{eqnarray}
A five-dimensional rotating EMCS black hole solution \cite{Reimers:2016czc}, in 
Boyer-Lindquist coordinates ($t, r, \theta, \phi, \psi$), can be expressed by the metric
\begin{eqnarray} \label{metric}
ds^2 &= &- \frac{\rho^2 dt^2 + 2q\nu dt}{\rho^2} + \frac{2q \nu \omega}{\rho^2} 
+ \frac{\mu \rho^2 - q^2}{\rho^4} \left(dt - \omega \right)^2 \nonumber \\
&& +\frac{\rho^2 dx^2}{4\Delta} + \rho^2 d\theta^2 + \left(x+a^2 \right) \sin^2 \theta d\phi^2 
\nonumber \\ 
& & + \left(x+b^2 \right) \cos^2 \theta d\psi^2,
\end{eqnarray}
where the gauge potential for the metric (\ref{metric}) can be expressed as
\begin{equation}\label{eq:gaugefield}
A_\mu dx^\mu = \frac{\sqrt{3} \, q}{\rho^2} \left(dt - \omega \right)
\end{equation}
and
\begin{eqnarray}
\Delta &=& \left(x+a^2 \right)\left(x+b^2 \right) + q^2 +2 abq - \mu x, \nonumber\\
\rho^2 &=& x + a^2 \cos^2 \theta + b^2 \sin^2 \theta, \nonumber\\
\nu &=& b \sin^2 \theta \, d\phi + a \cos^2 \theta d\psi, \nonumber\\
\omega &=& a \sin^2 \theta d\phi + b \cos^2 \theta  d\psi,
\end{eqnarray}
where $\mu$ is related to the black hole mass, $q$ is the charge and $a, b$ are the two 
different angular momenta of the black hole. Furthermore, the non-zero components of the 
metric (\ref{metric}) can be expressed as
\begin{eqnarray}\label{components}
g_{tt} &=& \frac{ \rho^2(\mu-\rho^2) -q^2 }{\rho^4},\nonumber\\
g_{t\phi} &=& -\frac{a (\mu \rho^2-q^2) + b q\rho^2\sin^2\theta}{\rho^4},\nonumber\\
g_{t\psi} &=& -\frac{b (\mu \rho^2-q^2) + a q\rho^2 \cos^2\theta }{\rho^4 },\nonumber\\
g_{\phi \psi} &=& \frac{[ ab(\mu \rho^2 -q^2) + (a^2+b^2) q \rho^2] 
\sin^2\theta \cos^2\theta}{\rho^4},\nonumber\\
g_{xx} &=& \frac{\rho^2}{\Delta},\qquad g_{\theta \theta}={\rho^2}, \nonumber\\
g_{\phi \phi} &=&  {(x+a^2)\sin^2\theta} + \frac{ a[a(\mu \rho^2 -q^2) + 2 bq \rho^2]
\sin^4\theta}{\rho^4 },\nonumber\\
g_{\psi \psi}  &=& {(x+b^2)\cos^2\theta} + \frac{ b[b(\mu \rho^2 -q^2) + 2 aq \rho^2]
\cos^4\theta}{\rho^4 }.
\end{eqnarray}
The radial coordinate has been changed to a new radial coordinate $x$ via $x=r^2$. One can 
check that, for $q=0$, the five-dimensional EMCS black hole reduces to the five-dimensional 
Myers-Perry black hole, which is analyzed in \cite{Myers:1986un,Papnoi:2014}, and also in 
addition if $a=0$, it reduces to five-dimensional Schwarschild-Tangherlini black hole 
\cite{Tangherlini:1963bw}. It may be noted that the metric (\ref{metric}) is independent of 
the coordinates ($t, \phi, \psi$), and hence it admits three Killing vectors given by 
\begin{eqnarray}
\ell &=& \frac{\partial}{\partial t} + \Omega_a\, \frac{\partial}{\partial \phi} 
    + \Omega_b \, \frac{\partial}{\partial \psi},
\end{eqnarray}
and these Killing vectors become null at the event horizon. The angular velocities for the 
metric (\ref{metric}) can be defined as
\begin{eqnarray}
\Omega_a &=& \frac{a(x^{H}_{+} + b^2)+ b q}{(x^{H}_{+} + a^2)(x^{H}_{+} +b^2)  + ab q},
\nonumber\\
\Omega_b &=& \frac{b(x^{H}_{+} + a^2)+ a q}{(x^{H}_{+} +a^2)(x^{H}_{+} +b^2)  + ab q},
\end{eqnarray}
where $x^{H}_{+}$ denotes the event horizon of five-dimensional EMCS black hole. The surface 
gravity of the EMCS black hole (\ref{metric}) takes the following form \cite{Chong:2005hr}:
\begin{eqnarray}\label{surface}
\kappa &=& \frac{(x^{H}_{+})^2-(ab + q)^2}
{\sqrt{x^{H}_{+}} [(x^{H}_{+} +a^2)(x^{H}_{+} +b^2) + abq]}.
\end{eqnarray} 
The five-dimensional EMCS black hole obeys the first law of thermodynamics and with the help 
of surface gravity ({\ref{surface}}), the  Hawking temperature of the black hole can easily 
be calculated via  $T=\kappa/(2\pi)$. The entropy of the black hole is given 
\cite{Chong:2005hr} by    
\begin{eqnarray}
S &=& \frac{\pi^2 [(x^{H}_{+}+a^2)(x^{H}_{+} + b^2) +a b q]}{2\sqrt{x^{H}_{+}}},
\end{eqnarray}
when $a=b=0=q$, it reduces to
\begin{eqnarray}
S &=& \frac{\pi^2 {(x^{H}_{+}})^{3/2}}{2}.
\end{eqnarray}
The Komar integral reads
\begin{equation}
J = \frac{1}{16\pi}\int_{S^3}  {*dK},
\end{equation}
where $K=\partial/\partial\phi$ or $K=\partial/\partial\psi$, yielding
\begin{eqnarray}
J_a = \frac{\pi(\mu a + qb)}{4}, \quad J_b = \frac{\pi(\mu b + qa)}{4}.
\end{eqnarray}
The electric charge can be calculated by the Gaussian integral
\begin{equation}
Q=\frac{1}{16\pi}\int_{S^3}( {*F} -F\wedge A/\sqrt3),
\end{equation}
which gives
\begin{eqnarray}
Q &=& \frac{\sqrt3\, \pi q}{4}.
\end{eqnarray}
As is well known the five-dimensional EMCS black hole follows the first law of thermodynamics 
so the conserved mass or the energy can be calculated by integrating   
\begin{equation}\label{de}
dE=TdS + \Omega_a dJ_a + \Omega_b dJ_b + \Phi dQ,
\end{equation}
where $\Phi$ is the electrostatic potential. Integrating Eq.~(\ref{de}), we obtain
\begin{eqnarray}
E &=& \frac{3\pi \mu}{8},
\end{eqnarray}
which represents the conserved energy of the five-dimensional EMCS black hole. 
Interestingly, the determinant is same as the uncharged case 
$\sqrt{- \det g} = \rho^2 \sin \theta \cos \theta /2$.
\subsection{Horizons and ergosphere}
Our aim is to discuss the effect of charge $q$ on the horizons and also on the ergosphere. It 
can be seen that the metric (\ref{metric}) is singular at $\rho^2=0$ and $\Delta=0$. Note that 
$\rho^2=0$ is a physical singularity and $\Delta = 0$ gives a coordinate singularity which 
defines the horizons of the metric (\ref{metric}). It turns out that $\Delta=0$ admits two 
roots \cite{Reimers:2016czc}
\begin{equation}\label{horizons}
x^{H}_{\pm} = \frac{1}{2} \left(\mu - a^2 - b^2 \pm \sqrt{\left(\mu - a^2 - b^2 \right)^2 
- 4\left(ab+q \right)^2} \right),
\end{equation}
which correspond to the five-dimensional EMCS black hole with two regular horizons, the event 
horizon ($x^H_{+}$) and the Cauchy horizon ($x^H_{-}$), when $(\mu -a^2-b^2)^2 > 4(ab+q)^2$ or 
$\mu > (a + b)^2 +2q $  \cite{Reimers:2016czc}. It represents an extremal black hole with 
degenerate horizons ($x^H_+=x^H_-$) when $ \mu = (a+b)^2 +2q$, and a naked singularity when 
$\mu < (a + b)^2 +2q $. If $q=0$, the Eq.~({\ref{horizons}) takes the following form:
\begin{equation}\label{horizons2}
x^{H}_{\pm} = \frac{1}{2} \left(\mu - a^2 - b^2 \pm \sqrt{\left(\mu - a^2 - b^2 \right)^2 
- 4a^2b^2} \right),
\end{equation}
where  the roots show the two horizons of the Myers-Perry spacetime 
\cite{Myers:1986un,Papnoi:2014}. If we consider the case when both spin parameters are equal 
$(a=b)$, then the Eq.~(\ref{horizons}) reduces to
\begin{equation}\label{equal}
x^{H}_{\pm} = \frac{1}{2} \left(\mu - 2a^2  \pm \sqrt{\left(\mu - 2a^2  \right)^2 
- 4\left(a^2+q \right)^2} \right).
\end{equation}
In this case the extremal black hole occurs at $\mu=2(2 a^2 +q)$, a naked singularity occurs 
at $\mu<2(2a^2+q)$, and a non-extremal black hole exists at $\mu>2(2a^2+q)$. Thus, the black 
hole charge affects the horizon structure and when $q=0$, Eq.~(\ref{equal}) reduces to
\begin{equation}\label{equalq}
x^{H}_{\pm} = \frac{1}{2} \left(\mu - 2a^2  \pm \sqrt{\left(\mu - 2a^2  \right)^2 
- 4a^4 } \right),
\end{equation}
which shows the two non degenerate horizons of five-dimensional Myers-Perry black hole 
\cite{Myers:1986un,Papnoi:2014} for equal rotation parameter. 
 \begin{figure*}
	 \includegraphics[width=0.45\linewidth]{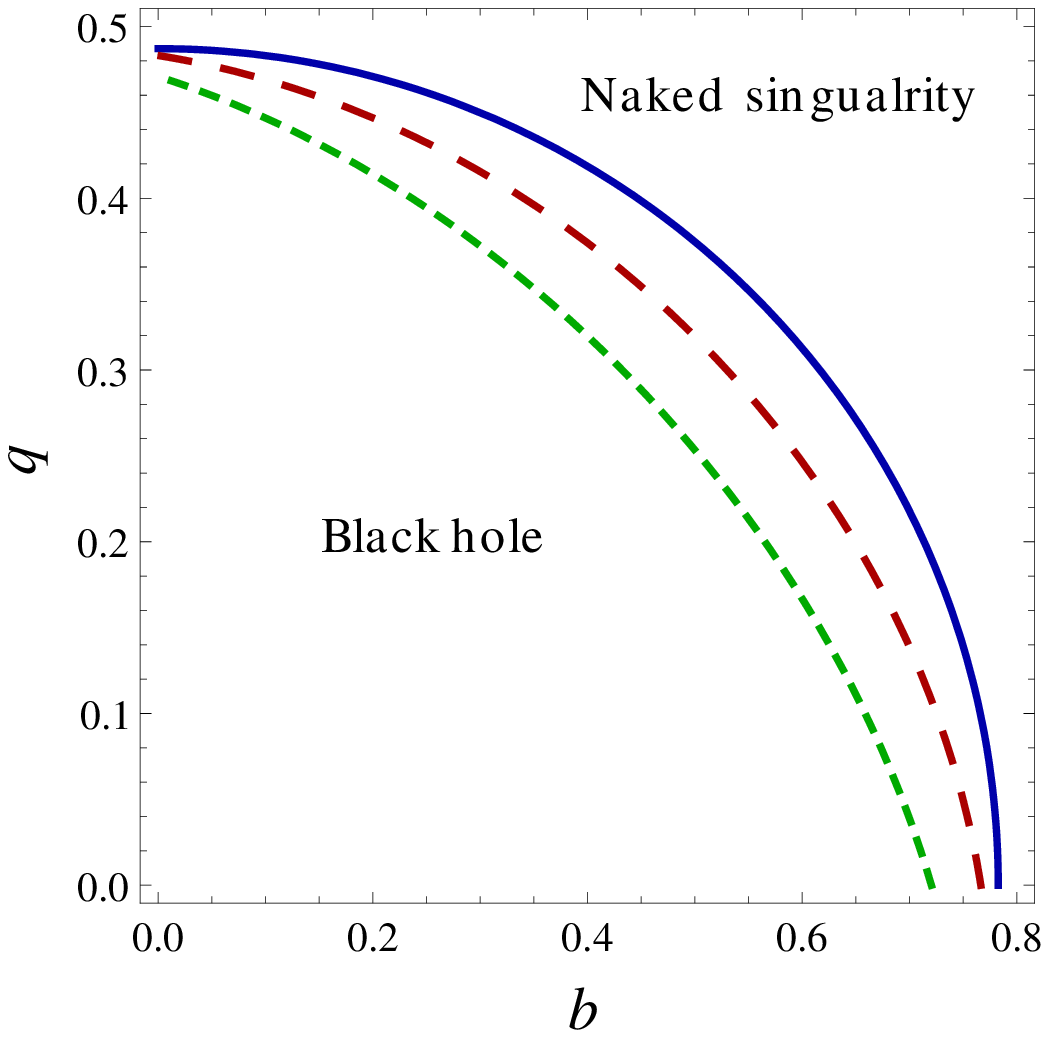}
	 \includegraphics[width=0.45\linewidth]{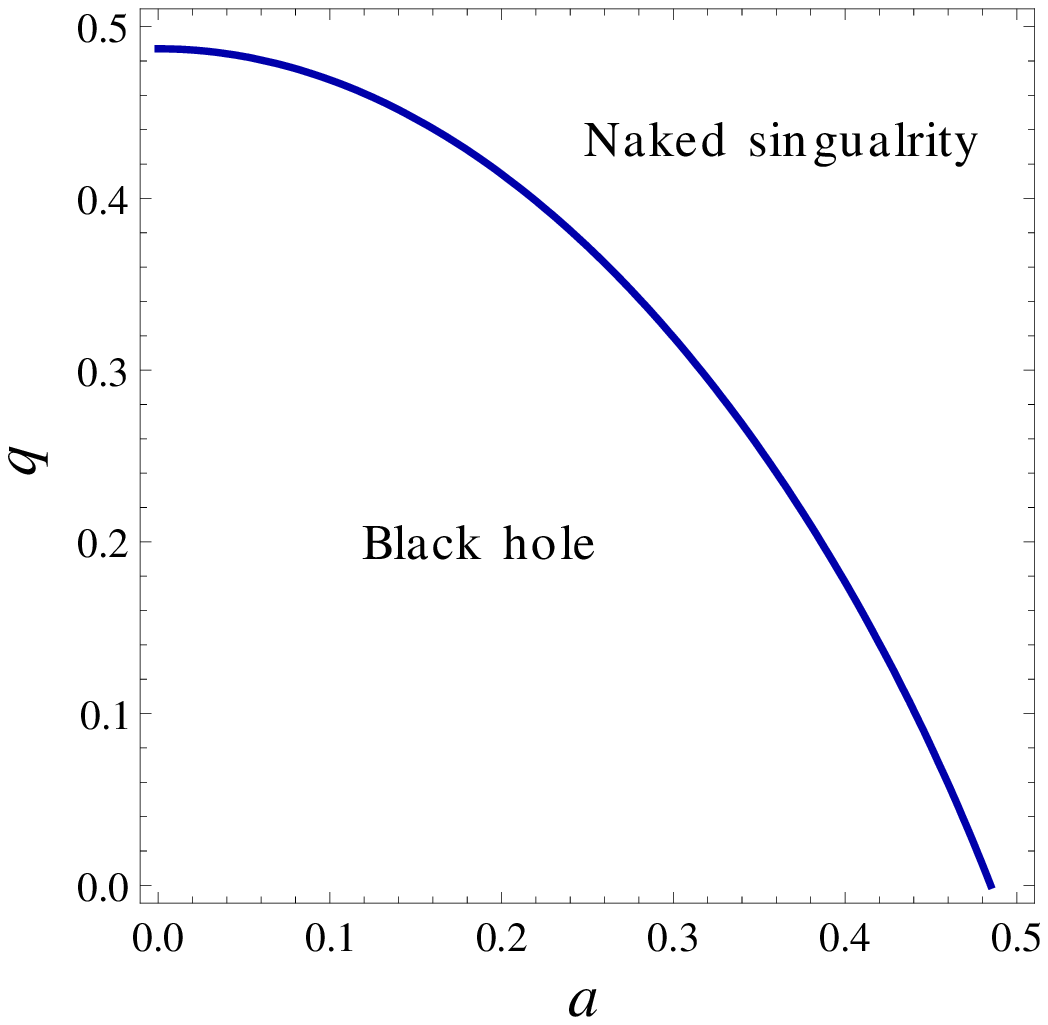}
\caption{\label{contourabq} Plot showing the dependence of charge $q$ with spin $a, b$. 
(Left) For $a \neq b$, $a=0$ (blue solid line), $a=0.1$ (red dashed line), $a=0.2$ 
(green dashed line). (Right) For $a=b$}
\end{figure*}
Next, we find the static limit surface where the time-like Killing vectors of the metric 
become null, i.e., $g_{tt}=0$, which leads to
\begin{equation}\label{slseq}
(x + a^2 \cos^2 \theta + b^2 \sin^2 \theta)^2 - \mu (x + a^2 \cos^2 \theta 
+ b^2 \sin^2 \theta)+q^2=0,
\end{equation}
which can be trivially solved \cite{Reimers:2016czc},
\begin{equation}\label{sls}
x^{sls}_{\pm}=\frac{1}{2} \left(\mu \pm \sqrt{\mu^2-4 q^2}\right)
-a^2 \cos^2 \theta-b^2 \sin^2 \theta,
\end{equation}
for equal rotation parameters $a=b$  \cite{Reimers:2016czc}, the roots can be defined as
\begin{equation}\label{sls1}
x^{sls}_{\pm}=\frac{1}{2} \left(\mu \pm \sqrt{\mu^2-4 q^2}\right)-a^2. 
\end{equation}
We can see that in the limit $q \rightarrow0$,
\begin{equation}
x^{sls}_{+} = \mu-a^2, \quad x^{sls}_{-} = a^2. 
\end{equation}
It can be seen that both surfaces, i.e., static limit surface and the horizons of the black 
hole, are not coinciding at the poles ($\theta =0, \pi/2$), and hence the ergosphere is 
totally different from the Kerr-Newman black hole where they do coincide at the poles (cf. 
Fig.~\ref{ergo}). The only possibility is when we have chosen \cite{Reimers:2016czc}
$$\theta = \arccos \left(\pm \sqrt{\frac{\mu}{a^2 + b^2}}\,\right), \quad 
\hbox{and} \quad q=-\frac{ab\mu}{a^2 + b^2}.$$
Hence, for the above values of $\theta$ and $q$, both surfaces of the metric (\ref{metric}) 
coincide. Figure~\ref{contourabq} shows the contour plots of $\Delta=0$, for the cases when 
$a\neq b$ and $a=b$. The contours depicted in Fig.~\ref{contourabq} indicate the boundary 
lines between a black hole region and the naked singularity. The coinciding roots arise on 
the colored line and there exist two roots inside the black hole region 
(cf. Fig.~\ref{contourabq}). It can also be seen  from Fig.~\ref{contourabq} that if there 
is no real root, then the region belongs to the naked singularity. 
\begin{figure*}
	\begin{tabular}{c c c c}
	 \includegraphics[width=0.25\linewidth]{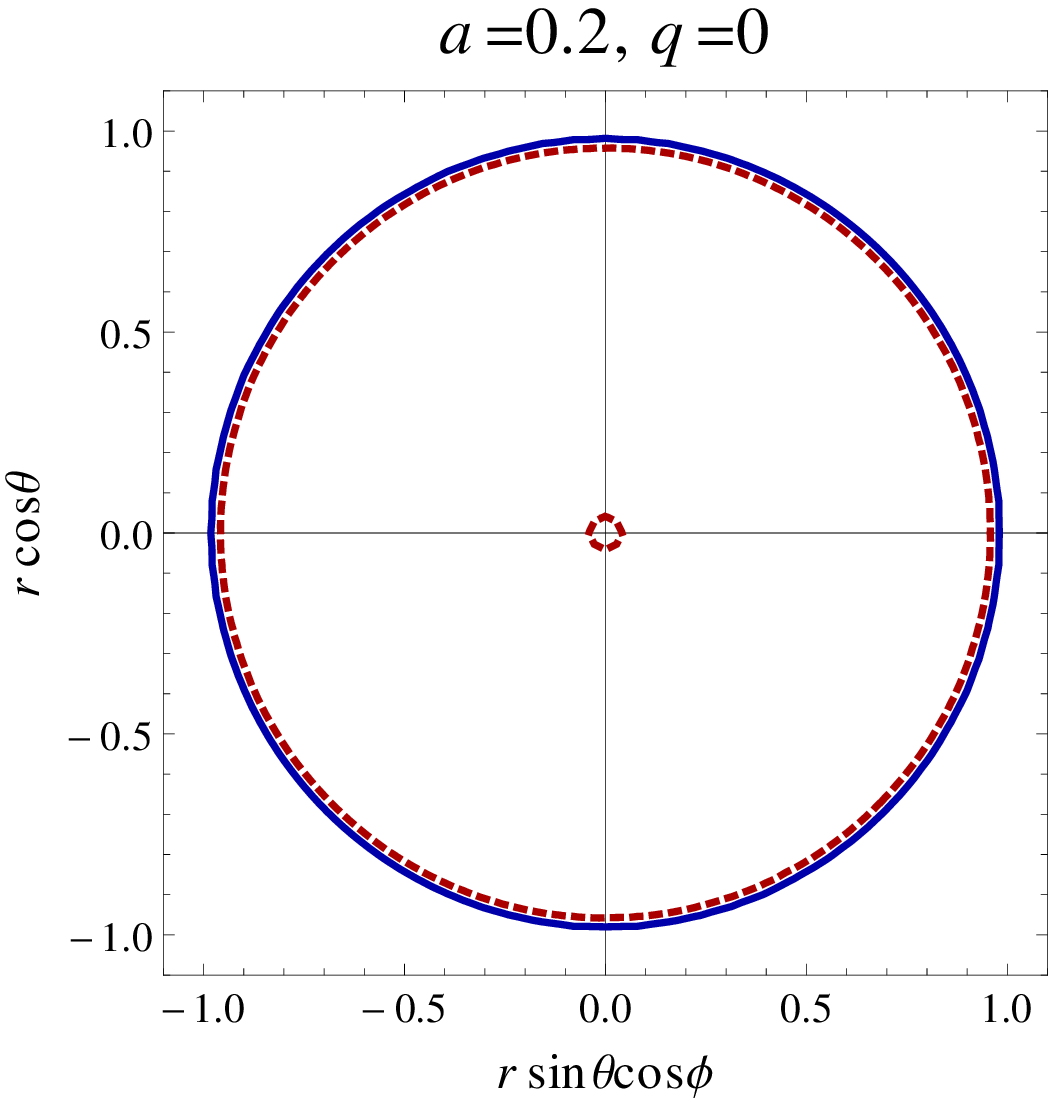}
	 \includegraphics[width=0.25\linewidth]{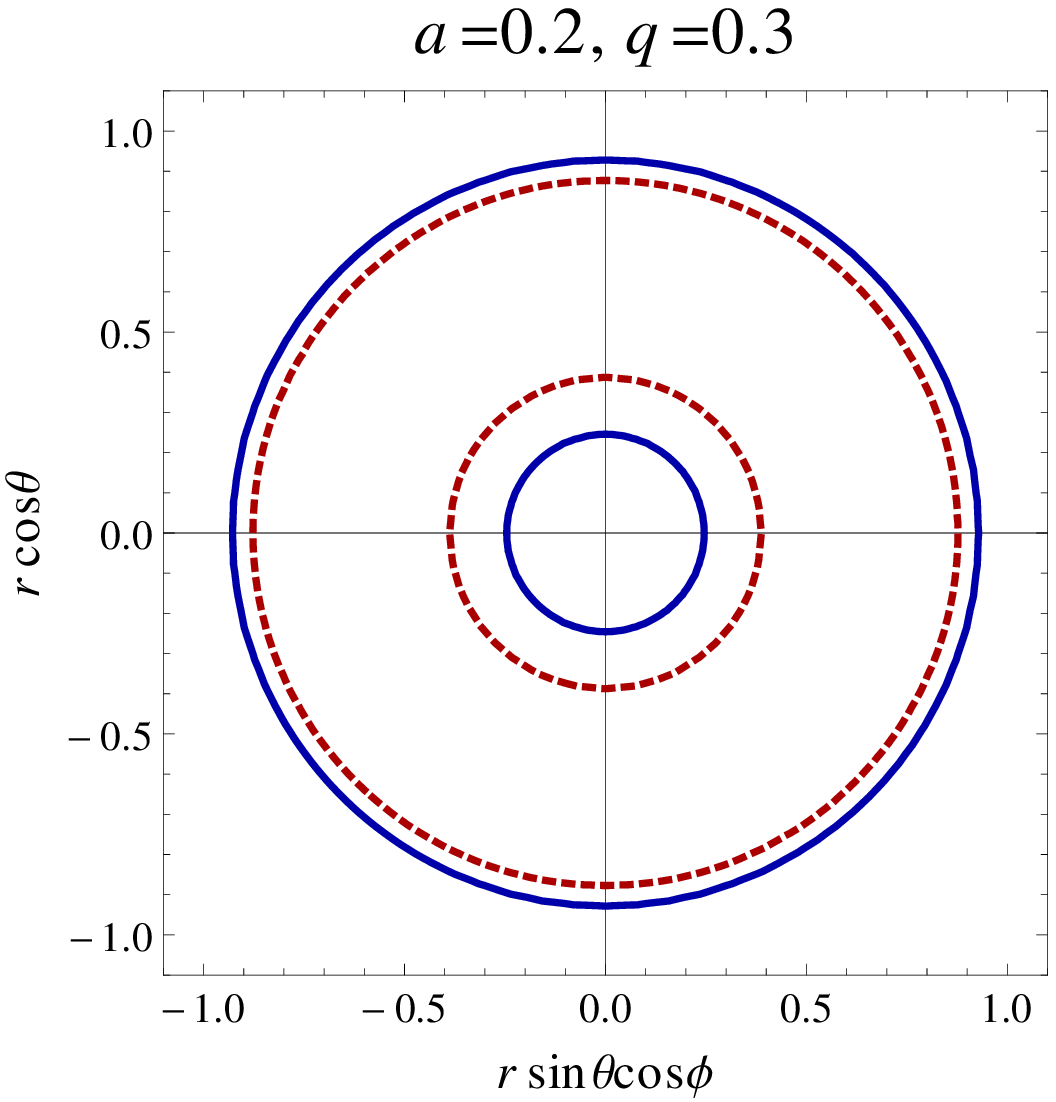}
	 \includegraphics[width=0.25\linewidth]{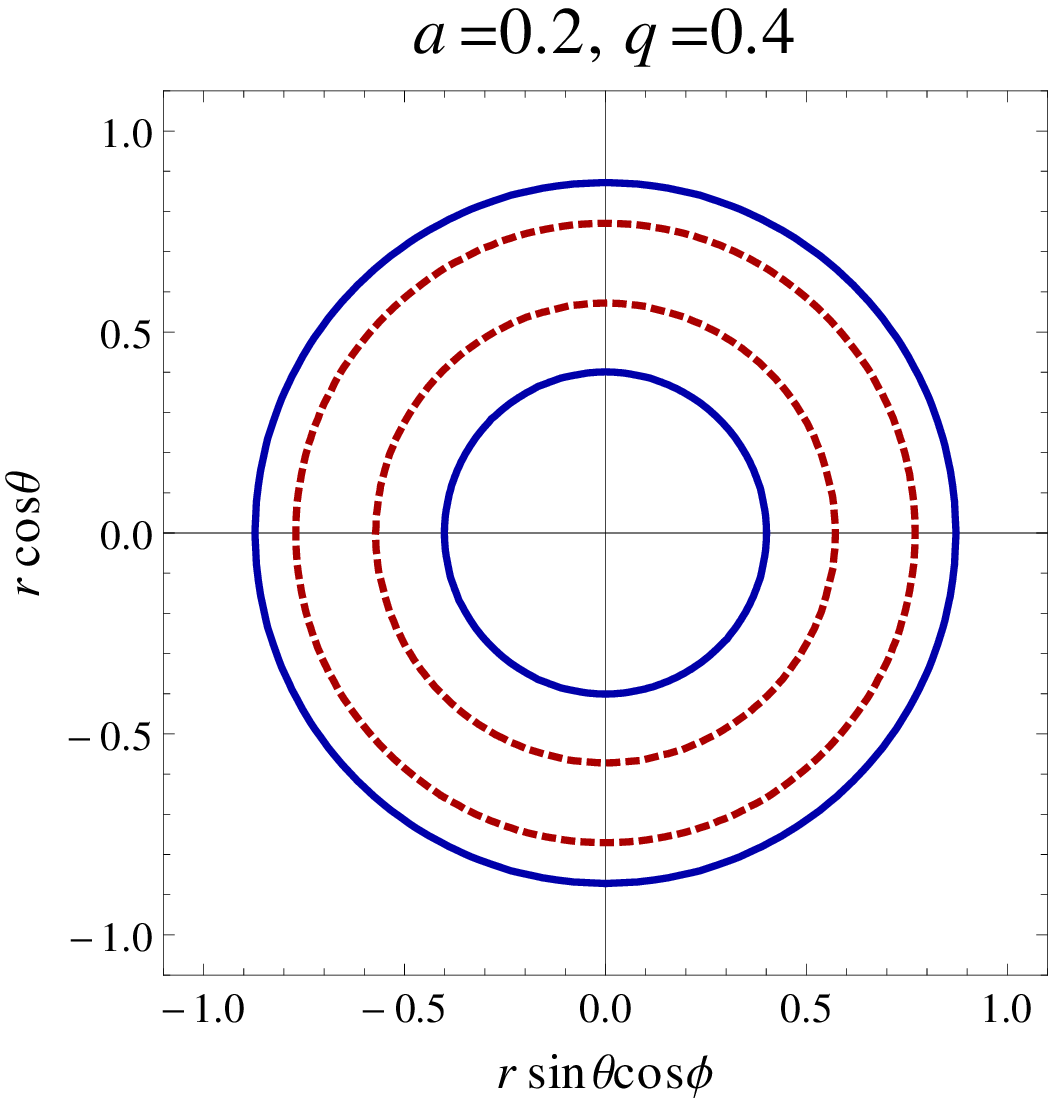}
	 \includegraphics[width=0.25\linewidth]{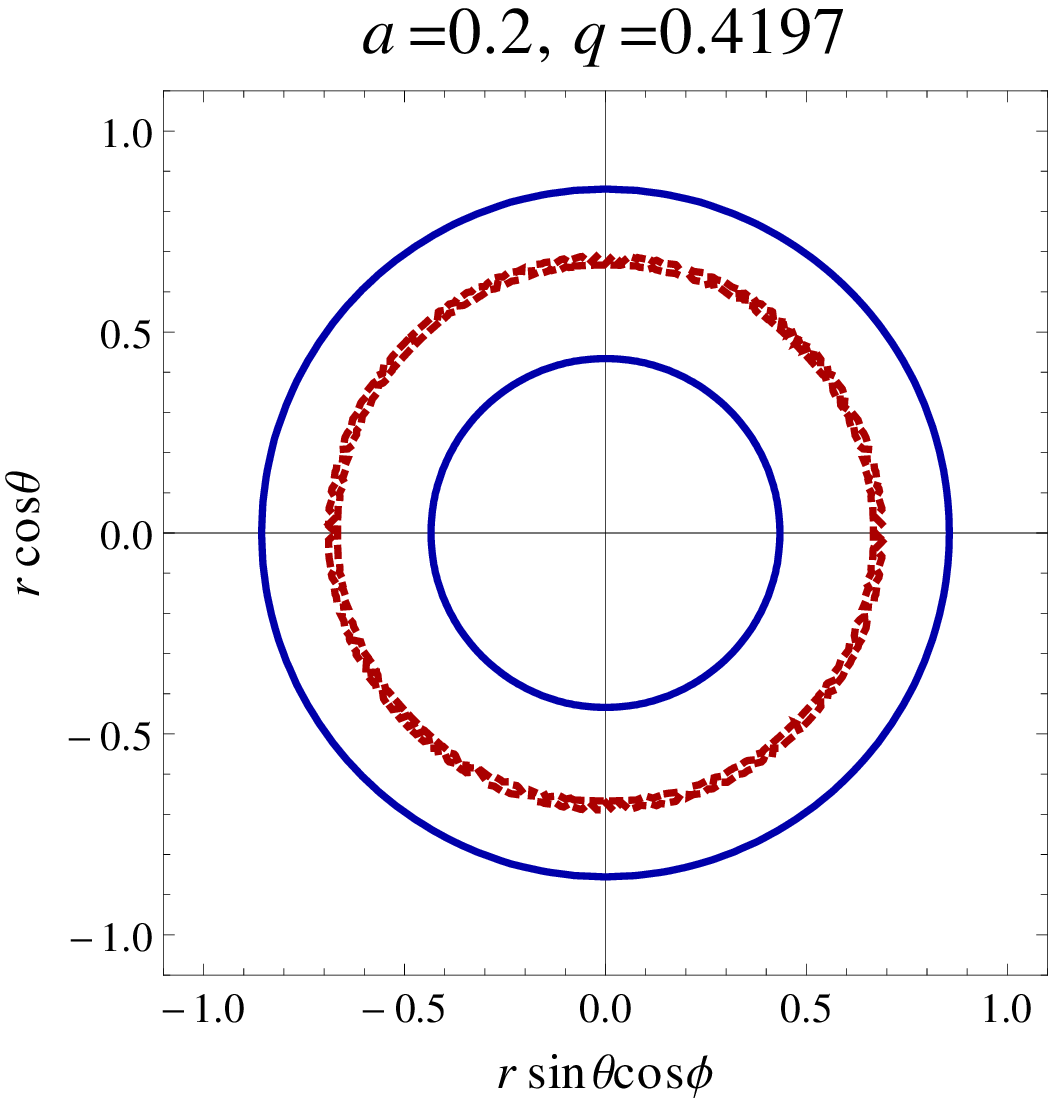}\\
	 \includegraphics[width=0.25\linewidth]{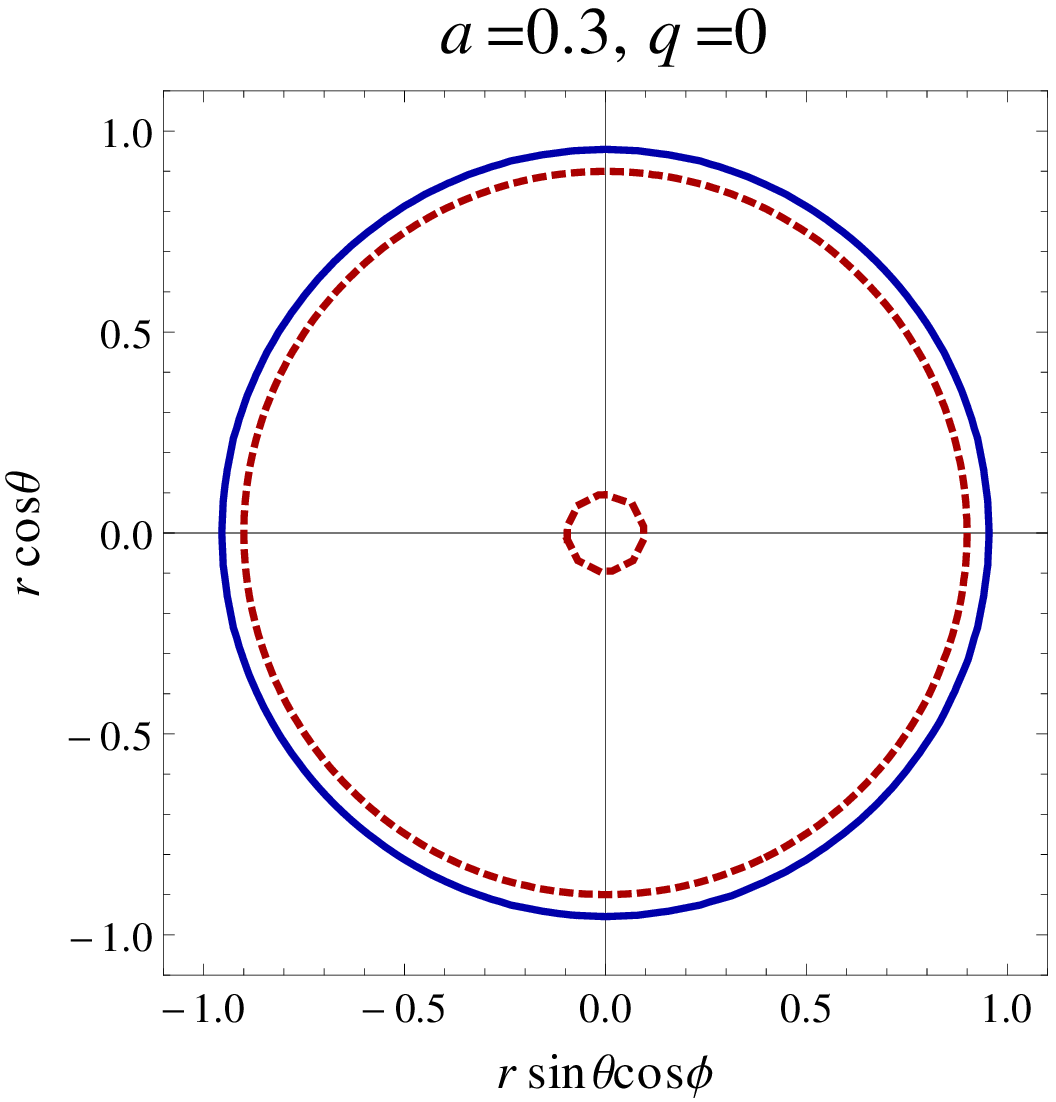}
	 \includegraphics[width=0.25\linewidth]{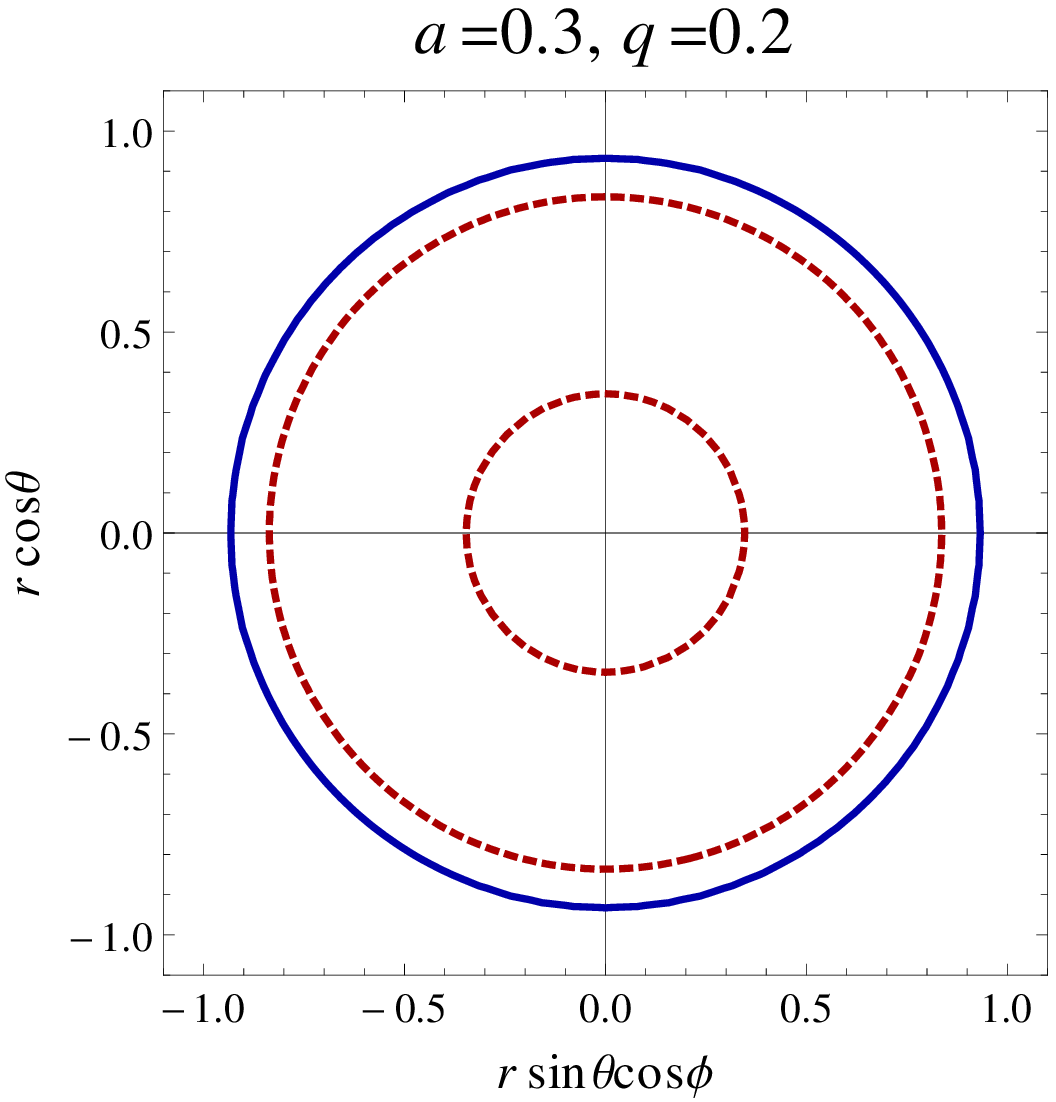}
	 \includegraphics[width=0.25\linewidth]{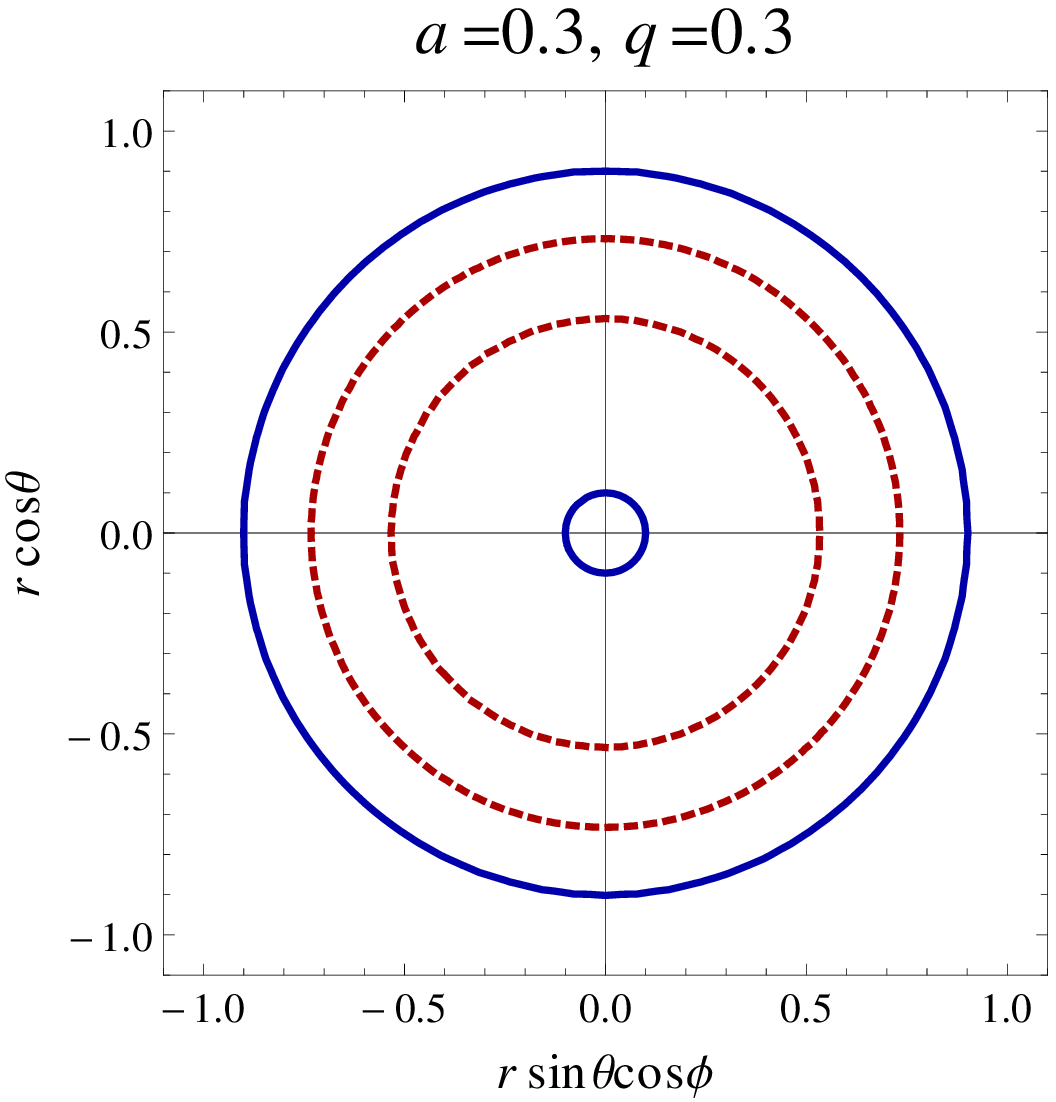}
	 \includegraphics[width=0.25\linewidth]{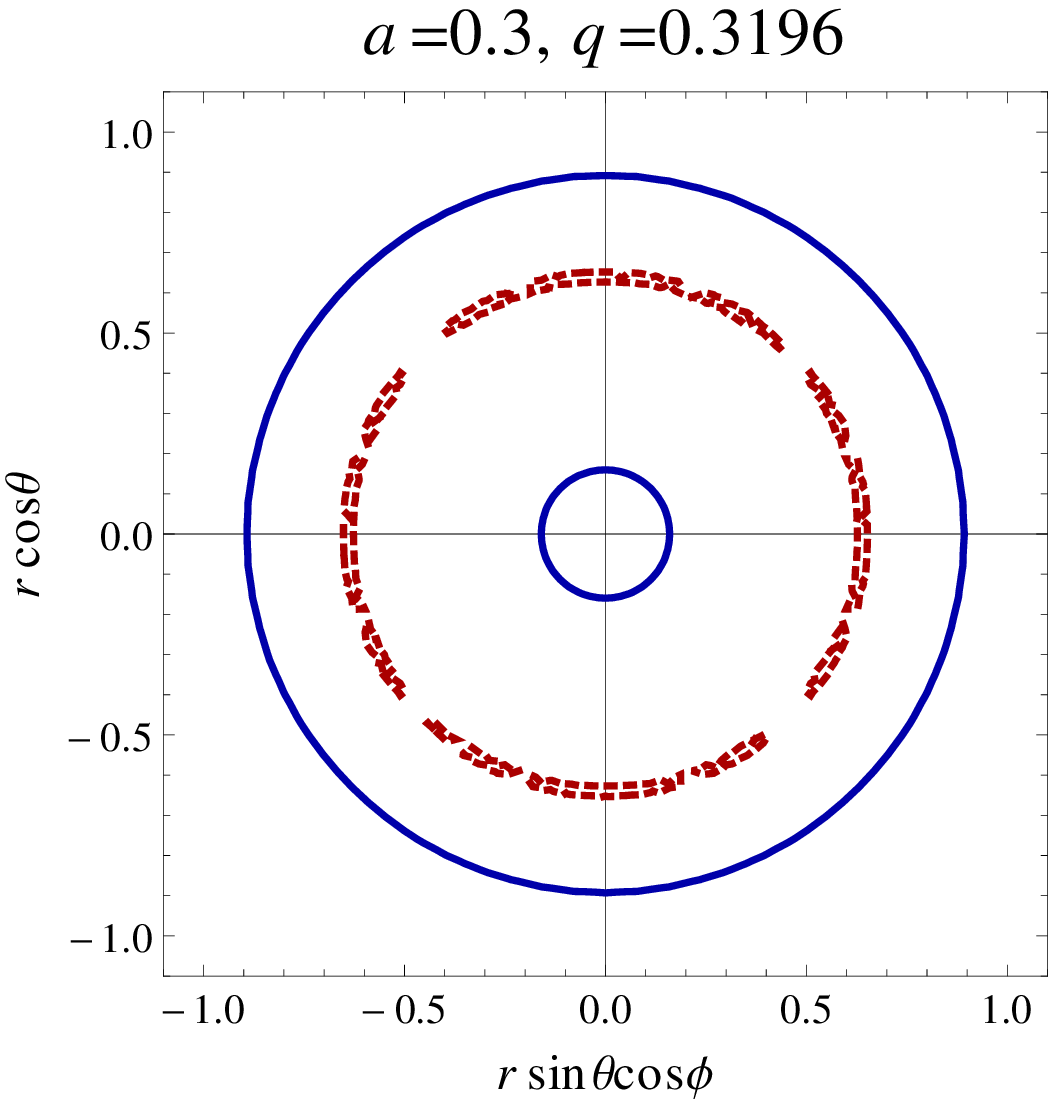}\\
	 \includegraphics[width=0.25\linewidth]{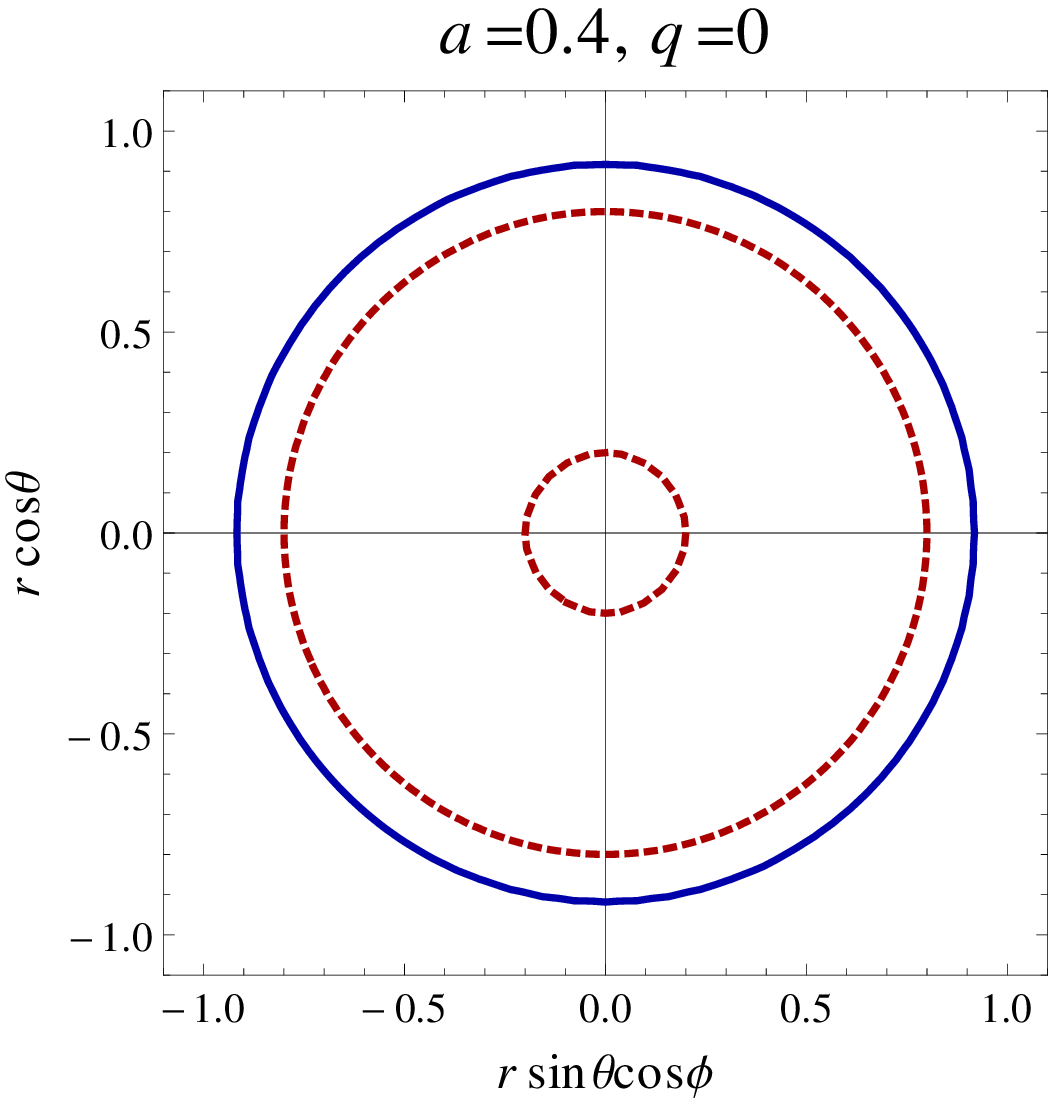}
	 \includegraphics[width=0.25\linewidth]{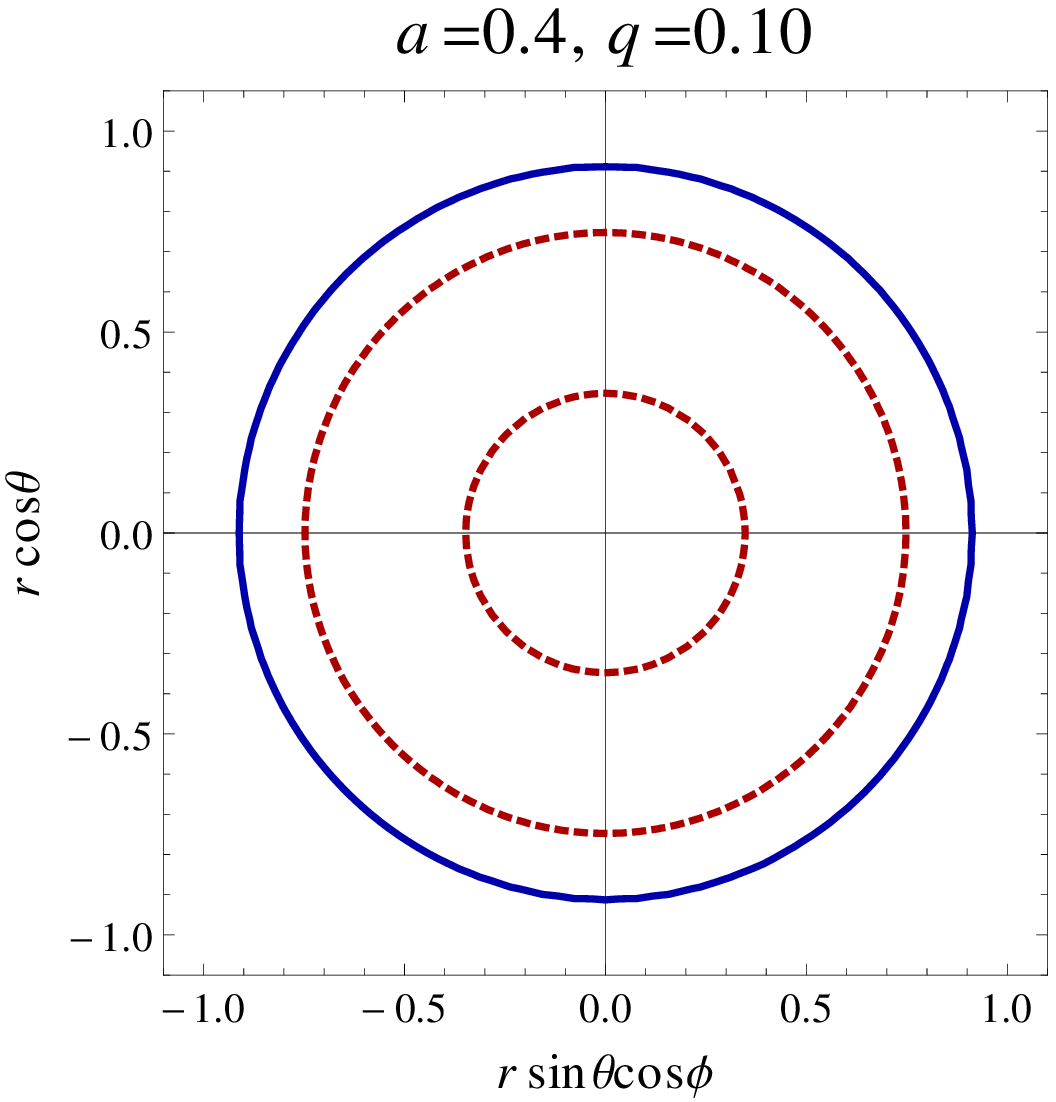}
	 \includegraphics[width=0.25\linewidth]{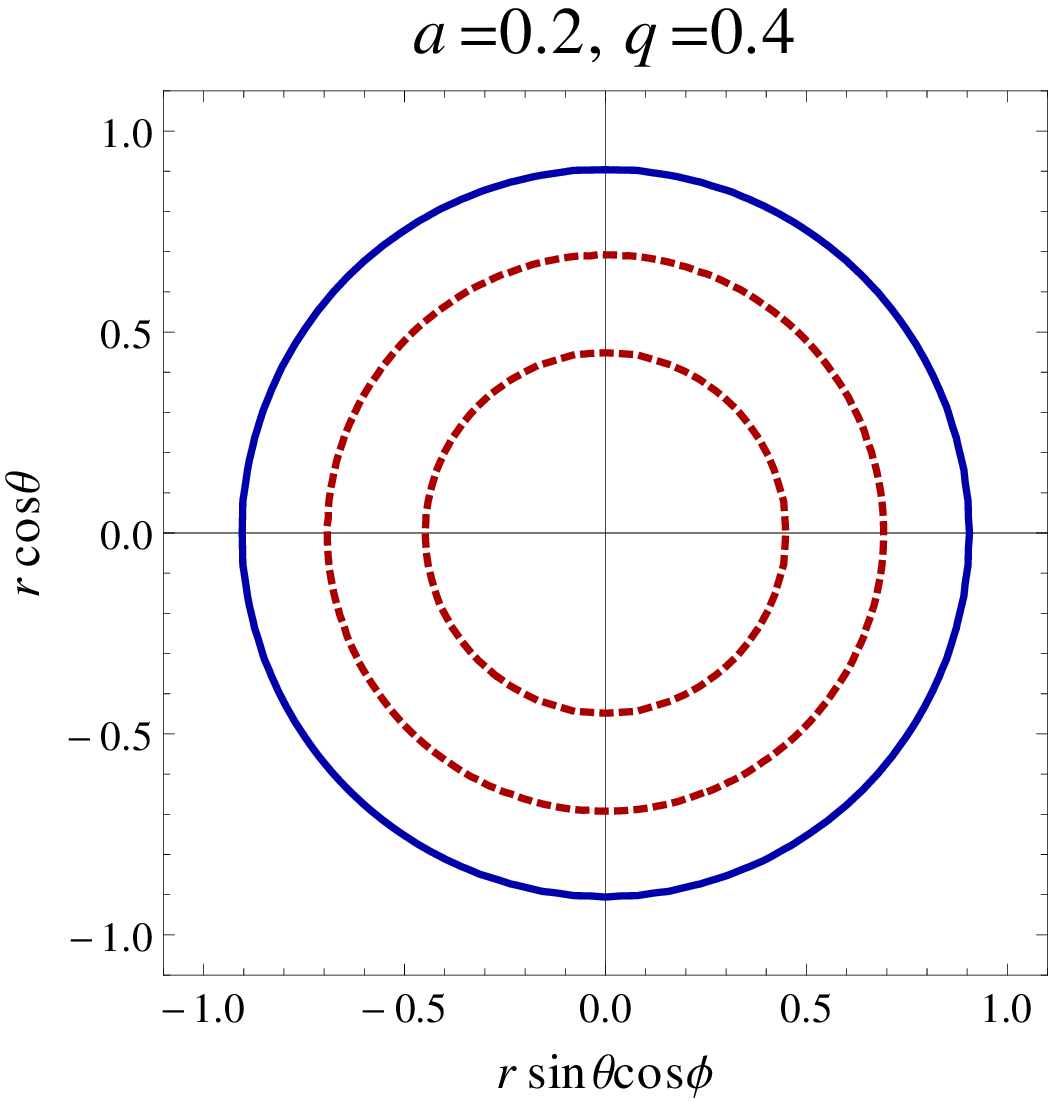}
	 \includegraphics[width=0.25\linewidth]{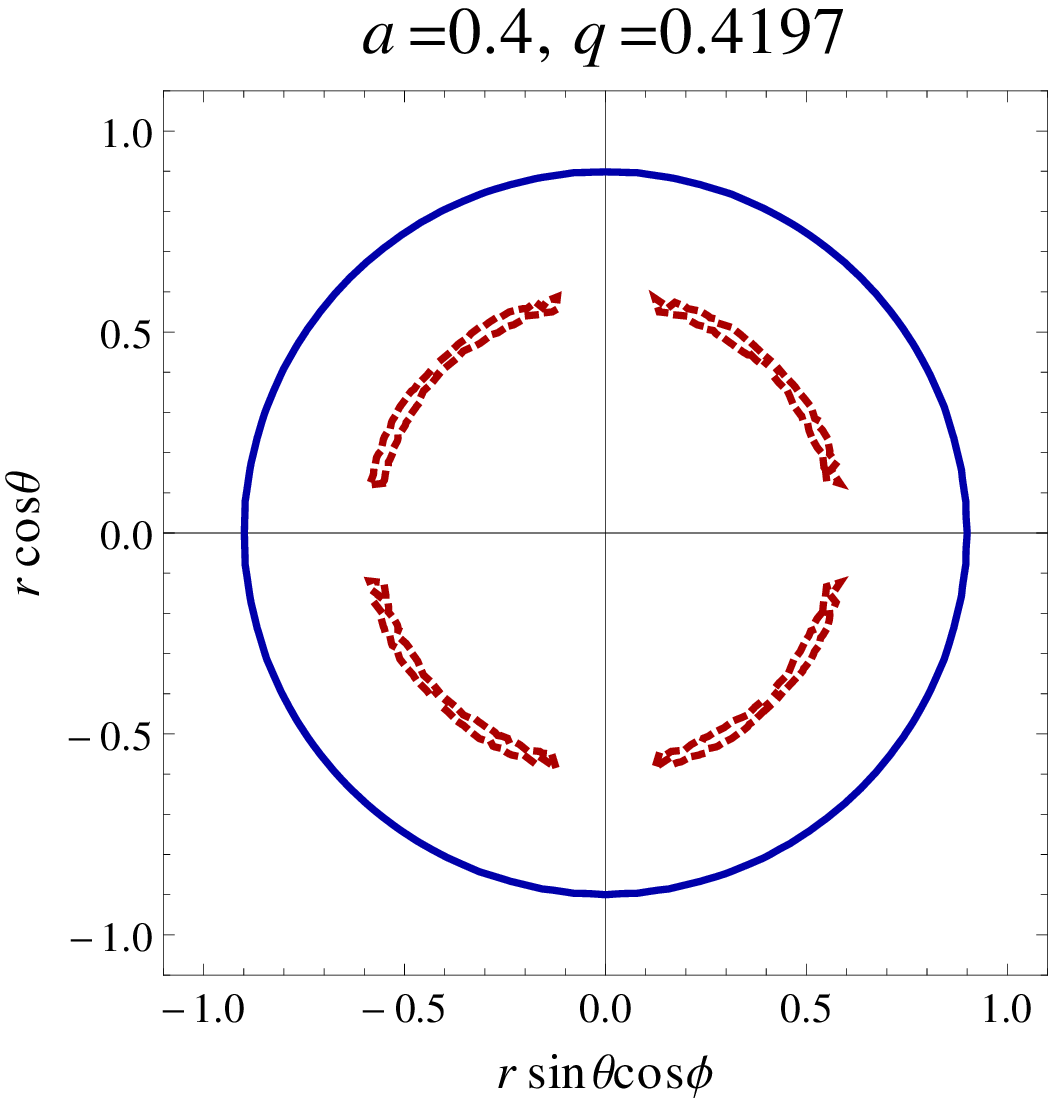}
	\end{tabular}
\caption{\label{ergo} The cross section of the event horizon, static limit surface, and 
ergoregion for different values of $q$ with a fixed value of rotation parameter $a$ for the 
five-dimensional EMCS black hole; the case $q=0$ refers to the Myers-Perry black hole. The 
blue line indicates static limit surface and the red one represents horizons}
\end{figure*} 
An ergoregion is a region outside the event horizon where the time-like Killing vectors behave 
like space-like. A particle can enter into the ergoregion and leave again, and it moves in the 
direction of spin of the black hole and has relevance for the energy extraction process 
\cite{Penrose}. We have plotted the ergoregion of the five-dimensional EMCS black hole in 
Fig.~\ref{ergo} for different values of $a$ and $q$, and we observe that there is an increase 
in the area of ergoregion when we increase the values of the parameter $q$ and $a$ 
(cf. Fig.~\ref{ergo}). 
\begin{figure*}
	\begin{tabular}{c c c c}
		\includegraphics[width=0.25\linewidth]{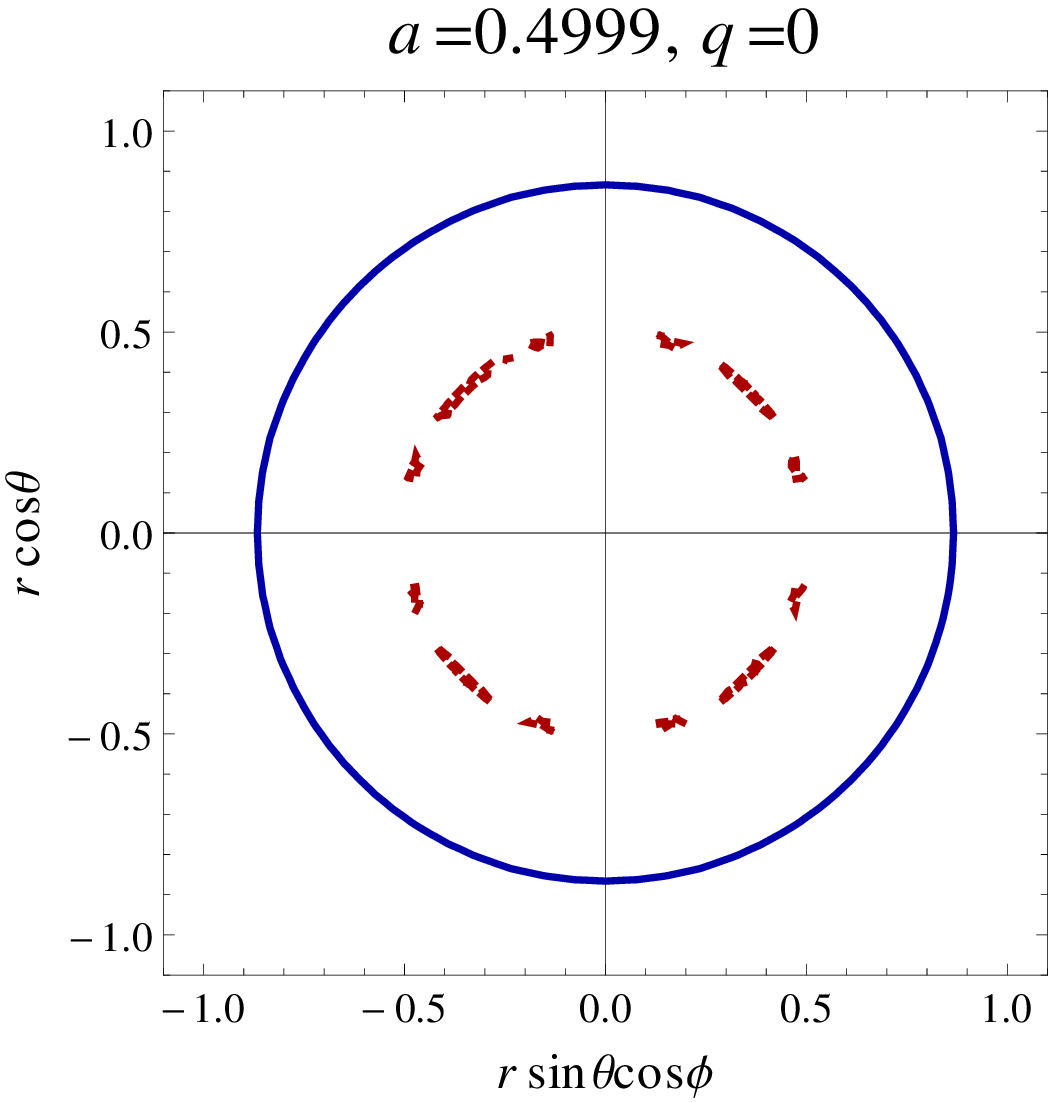}
		\includegraphics[width=0.25\linewidth]{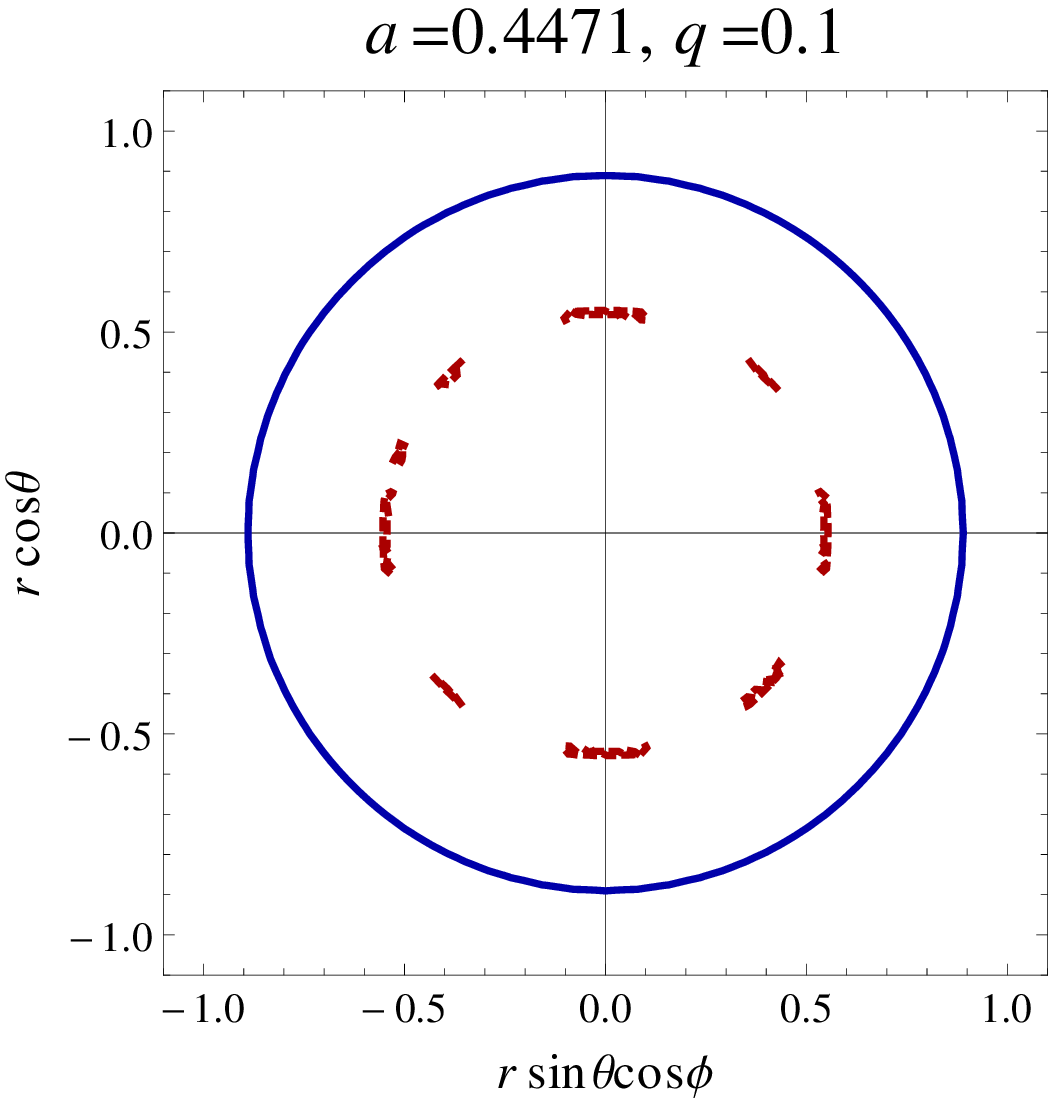}
		\includegraphics[width=0.25\linewidth]{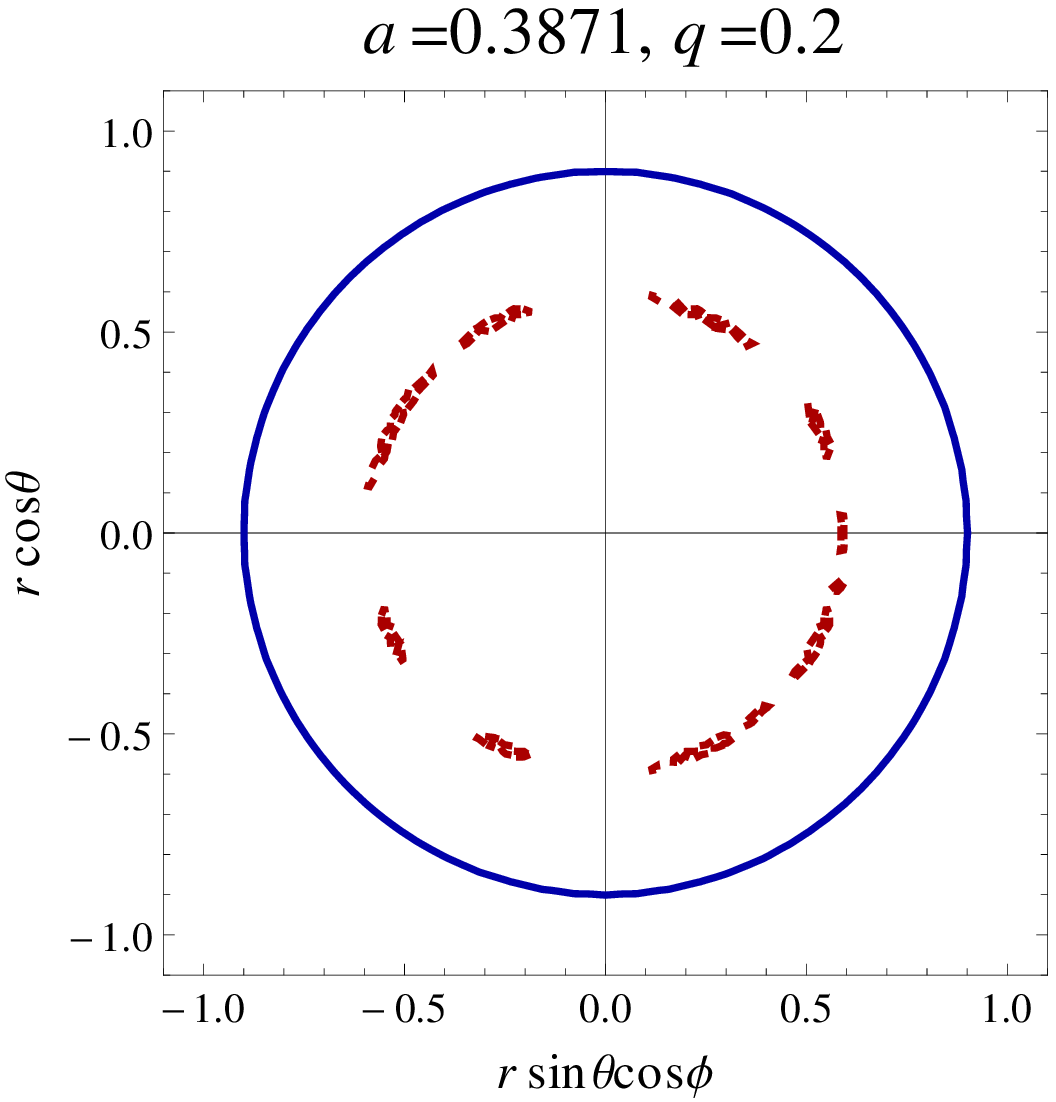}
		\includegraphics[width=0.25\linewidth]{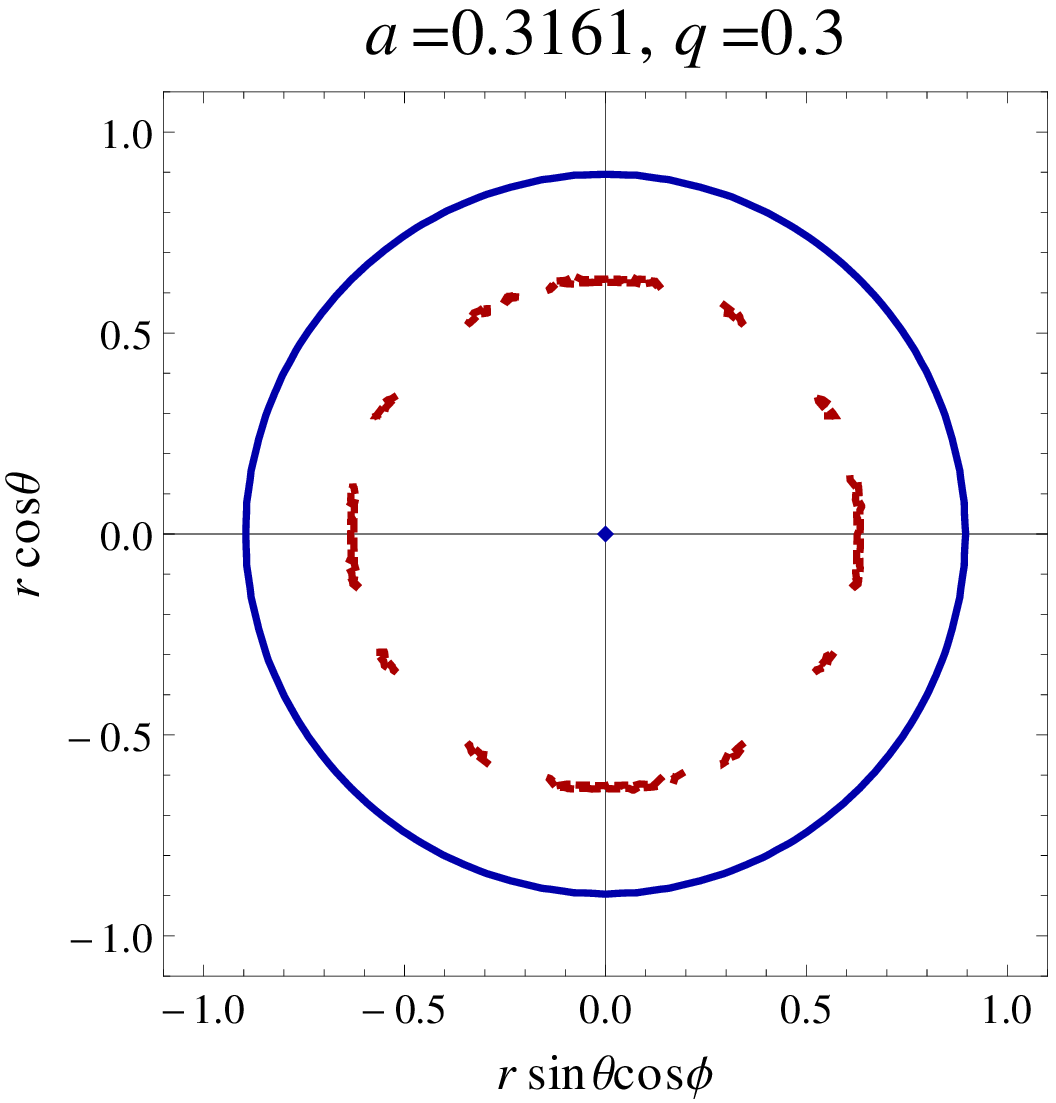}
	\end{tabular}
	\caption{\label{ergo1} The cross section of the event horizon, static limit surface, and 
	ergoregion for different values of $q$ with a fixed rotation parameter $a$ for the 
	extremal five-dimensional EMCS black hole when two horizons coincide. The blue lines 
	indicate static limit surface and the red ones represent horizons}
\end{figure*} 

In Kerr black hole case, we have the region between the event horizon and static limit surface 
known as the ergoregion, which can be seen from  Fig.~\ref{ergo}. In Fig.~\ref{ergo}, we 
explicitly show  the effect of parameter $q$ on the ergoregion, it shows that the area of 
ergoregion grows with rotation parameter $a$ as well as with charge $q$. Thus, for the faster 
rotating black holes the ergoregion is enlarged. Figure~\ref{ergo1} suggests that the 
ergoregion area is maximum for an extremal five-dimensional EMCS black hole.

\section{Particle motion}
\label{pm}
Different tests were proposed to discover signatures of extra dimensions in supermassive black 
holes since the gravitational field may be different from the standard one in the general 
relativity approach. In particular, gravitational lensing features are different for 
alternative gravity theories with extra dimensions and general relativity. The study of 
higher-dimensional black hole shadow may help us to understand how measurements of the sizes 
can put constraints on the parameters of a black hole in the spacetime with extra dimensions. 
Next, we would like to study the shadow of the EMCS black hole, which requires a complete 
study of particle motion around the black hole, and can be obtained by Hamilton-Jacobi 
formalism originally suggested by Carter \cite{Carter:1968rr}. The Hamilton-Jacobi equation 
\cite{Chandrasekhar} for the five-dimensional EMCS black hole (\ref{metric}) with the metric 
tensor $g_{\mu \nu}$ (\ref{components}) reads
\begin{equation}
\label{eq:HamJac}
- \frac{\partial S}{\partial \sigma} = \frac{1}{2} g^{\mu \nu}\frac{\partial S}
{\partial x^\mu} \frac{\partial S}{\partial x^\nu},
\end{equation}
where $\sigma$ is the affine parameter and $S$ is the Jacobian action with the following 
separable ansatz for the Jacobian action \cite{Papnoi:2014}:
\begin{equation}
\label{eq:Sansatz}
S = \frac{1}{2} {m{_0}^2} \sigma - Et + S_r(r) + S_\theta(\theta) + L_{\phi} \phi 
+ L_{\psi}\psi,
\end{equation} 
where $m_0$ is the mass of test particle, which is zero in case of photon, the conserved 
quantities $E$, and $L_{\phi}$, $L_{\psi}$ correspond to the energy and angular momentum of 
the particle, respectively. However, $S_r(r)$ and $S_{\theta}(\theta)$ are, respectively,
functions of $r$ and $\theta$ only. It can be seen, using Eqs.~\eqref{eq:Sansatz} and 
\eqref{eq:HamJac}, one can find the complete geodesic equations for the five-dimensional EMCS 
black hole \cite{Reimers:2016czc} :
\begin{eqnarray}
\label{eq:ut}
\rho^2 \frac{dt}{d\sigma} &=& E\rho^2+\frac{1}{\Delta}\Big[\left(\mu \zeta\gamma
-q^2(\zeta+b^2 \right)E 
\nonumber\\
&& + \left(a\gamma \mu +b\gamma q-aq^2\right)L_\phi +\left(\zeta b \mu +a\zeta q-bq^2\right)
L_\psi \Big], \\
\label{eq:uphi}
\rho^2 \frac{d\phi}{d\sigma} &=& \frac{L_{\phi}}{\sin^2\theta}-\frac{1}{\Delta}
\Big[\Big((a^2-b^2)\gamma + \mu b^2+2abq\Big)L_{\phi} \nonumber\\
 && + \Big(\mu ab+(a^2+b^2)q\Big)L_{\psi}+\Big(a\gamma \mu +b\gamma q-aq^2\Big)E\Big],\\
\label{eq:upsi}
\rho^2 \frac{d \psi}{d\sigma} &=& \frac{L_\psi}{\cos^2\theta}-\frac{1}{\Delta}
\Big[\left(-(a^2-b^2)\zeta + \mu a^2+2abq\right)L_\psi \nonumber\\ 
&& + \left(\zeta b\mu +a\zeta q-bq^2\right)E+\Big(\mu ab+(a^2+b^2)q\Big)L_\phi\Big],\\
\label{eq:ux}
\rho^2 \frac{dx}{d\sigma} &=& \pm \sqrt{\cal R}, \\
\label{eq:utheta}
\rho^2 \frac{d \theta}{d\sigma} &=& \pm \sqrt{\Theta},
\end{eqnarray}
where $\zeta = x+a^2$ and $\gamma = x+b^2$. In geodesic Eqs.~(\ref{eq:ux}) and 
(\ref{eq:utheta}), the terms $\cal {R}$ and $\Theta$ are given by
\begin{eqnarray}
\mathcal{R} &=& 4 \left(E^2 \Delta x - \Delta \mathcal{K} + \mathcal{E} + \mu \mathcal{M} 
+ 2q \mathcal{Q} - q^2 \mathcal{P}\right),  \\
\Theta &=& E^2 \left(a^2 \cos^2 \theta + b^2 \sin^2 \theta \right) + \mathcal{K}
-\frac{L_\phi^2}{\sin^2\theta}- \frac{L_\psi^2}{\cos^2 \theta},
\end{eqnarray}
where $\mathcal{K}$ is the Carter constant \cite{Carter:1968rr} and 
\begin{eqnarray}
\mathcal{E} &=& (a^2-b^2)(\gamma L_\phi^2 -\zeta L_\psi^2), \nonumber\\ 
\mathcal{Q} &=& ab \left(L_{\phi}^2 + L_{\psi}^2\right) +  (a^2+b^2) L_{\phi} L_{\psi} 
+ Eab \left(\frac{L_{\phi} \gamma}{a} + \frac{L_{\psi} \zeta}{b} \right), \nonumber\\ 
\mathcal{M} &=& \zeta \gamma E^2 + 2 a \gamma E L_\phi + 2 \zeta bEL_{\psi} + (b L_{\phi} 
+ aL_{\psi})^2, \nonumber\\ 
\mathcal{P} &=& 2a EL_{\phi} + 2b E L_{\psi} + (\zeta+b^2)E^2. \nonumber
\end{eqnarray}
These geodesic equations define the geometry of photon around the spacetime of 
five-dimensional EMCS black hole. For a particle that moving in the equatorial plane and to 
remain in equatorial plane, it is necessary that the Carter constant $\mathcal{K}$ must be 
zero \cite{Bardeen:1972rr}. One can recover the equations of motion of five-dimensional 
Myers-Perry black hole when $q=0$ \cite{Papnoi:2014}. In the presence of a bright object 
behind a black hole or for obtaining the boundary of the  black hole shadow, the study of 
radial equation. We can rewrite the radial equations of motion \cite{Wei:2013kza,Papnoi:2014}  
\begin{eqnarray}
\left(\frac{dx}{d\sigma}\right)^2 + V_{eff}  &=& 0
\end{eqnarray}
with the effective potential
\begin{eqnarray}\label{veff}
V_{eff}&=& -\frac{4}{\rho^2} \Big[E^2 \Delta x - \Delta \mathcal{K} + \mathcal{E} + \mu\mathcal{M} 
+ 2q\mathcal{Q} - q^2 \mathcal{P}\Big].
\end{eqnarray}
When $q=0$,  Eq.~(\ref{veff}) reduces to
\begin{eqnarray}\label{veff1}
V_{eff}&=& -\frac{4}{\rho^2} \Big[E^2 \Delta x - \Delta \mathcal{K} + \mathcal{E} 
+ \mu\mathcal{M}   \Big],
\end{eqnarray}
which is the effective potential of the Myers-Perry black holes \cite{Frolov:2003en}. The 
apparent shape of the five-dimensional EMCS black hole can be obtained by a study of the 
photon orbits. 

The impact parameters characterizing the photon orbits around the black hole can be defined 
in terms of the constants of motion, i.e., $\xi_{1}=L_{\phi}/E$, $\xi_{2}=L_{\psi}/E$ and 
$\eta={\cal K}/E^2$. Therefore, the expression for $\cal R$ takes the form  
\begin{eqnarray}
\mathcal{R} &=&4 E^2 \left(\Delta x - \Delta \eta +\mu \mathcal{J} 
+ 2q\mathcal{O} - q^2 \mathcal{S}\right),
\end{eqnarray}
where $\mathcal{J} = \zeta \gamma + 2 a (\gamma \xi_{1} + \zeta \xi_{2}) + a^2(\xi_{1} 
+ \xi_{2})^2$, $\mathcal{O} = a^2 (\xi_{1}^2 + \xi_{2}^2) + 2a^2 \xi_{1} \xi_{2} 
+ a \left(\xi_{1} \gamma + \xi_{2} \zeta \right)$, and $\mathcal{S} = 2a (\xi_{1} + \xi_{2}) 
+ (\zeta+a^2)$. Henceforth, we assume that the two rotation parameters $a=b$. 

The radial motions of the photons are essential for determining the shadow of the 
five-dimensional EMCS black hole. The spherical photon orbits, i.e., geodesics that 
stay on a sphere $r$=constant, define the apparent shape of the black hole. The photons come 
from infinity and approach  a turning point with zero radial velocity, which corresponds to 
an unstable circular orbit determined by
\begin{equation}\label{condition}
V_{eff}=0 \quad \text{and} \quad \frac{d V_{eff}}{d x}  =0, 
\quad \text{or} \quad \mathcal{R}=0 \quad \text{and} \quad \frac{d\mathcal{R}}{dx}=0.
\end{equation}
By Eqs.~(\ref{veff}) and (\ref{condition}), as in the Kerr case \cite{Hioki:2009na}, 
one can obtain the parameters $\eta$ and the sum of the parameters $\xi_1$, $\xi_2$, which 
read
\begin{eqnarray}\label{eta}
\eta &=& \frac{1}{a(2q+1)(-1+2 a^2+2 x)^2}\Big[2a^7 (1+2 q) +a^5 [1+2 q (3+3 q+10 x)+10 x]
\nonumber\\
&+& 2 a^3 \left[q (1+q) (1+4 q)+2 (-1+(-1+q) q) x+7 (1+2 q) x^2\right]\nonumber\\
&-& 2 (1+q) \sqrt{a^2 \left[q^2+a^2 (2+4 q)+2 x+4 q x\right] \left[a^4+q^2+(-1+x) x
+2 a^2 (q+x)\right]^2}
\nonumber\\
&+& a \left[2 q^3 (1+q)+4 q^2 (1+q) x-(5+2 q (5+q)) x^2+6 (1+2 q) x^3\right]\Big],
\end{eqnarray}
and
\begin{eqnarray}\label{sumxi}
\xi_1 +\xi_2 &=& \frac{1}{a^2(2q +1)(-1+2 a^2+2 x)}\Big[-a^5 (1+q)+a^3 
\left[1+3 q+4 q^2-2 (1+q) x\right] 
\nonumber\\
&+& a \left[q^3+2 q^2 x-(1+q) x^2\right]\nonumber\\
&-& \sqrt{a^2 \left[q^2+a^2 (2+4 q)+2 x+4 q x\right] \left[a^4+q^2+(-1+x) x+2 a^2 (q+x)
\right]^2}\Big].
\end{eqnarray}
The impact parameters $\xi_1$, $\xi_2$ and $\eta$ for the photon orbits around 
five-dimensional EMCS black hole determine the contour of the shadow \cite{Papnoi:2014}. Now 
we consider the case when $\theta=\pi /2$, $L_{\psi}=0$, which implies $\xi_2=0$; therefore 
from Eq.~(\ref{sumxi}), we obtain
\begin{eqnarray}
\xi_1  &=& \frac{1}{a^2(2q +1)(-1+2 a^2+2 x)}\Big[-a^5 (1+q)+a^3 \left[1+3 q+4 q^2-2 (1+q) x
\right] \nonumber\\
&+& a \left[q^3+2 q^2 x-(1+q) x^2\right]\nonumber\\
&-& \sqrt{a^2 \left[q^2+a^2 (2+4 q)+2 x+4 q x\right] \left[a^4+q^2+(-1+x) x+2 a^2 (q+x)
\right]^2}\Big],
\end{eqnarray}
and for $\theta=0$, $L_{\phi}=0$, which implies $\xi_1=0$, thus
\begin{eqnarray}
\xi_2 &=& \frac{1}{a^2(2q +1)(-1+2 a^2+2 x)}\Big[-a^5 (1+q)+a^3 \left[1+3 q+4 q^2-2 (1+q) x
\right]
\nonumber\\
&+& a \left[q^3+2 q^2 x-(1+q) x^2\right]\nonumber\\
&-& \sqrt{a^2 \left[q^2+a^2 (2+4 q)+2 x+4 q x\right] \left[a^4+q^2+(-1+x) x+2 a^2 (q+x)
\right]^2}\Big].
\end{eqnarray}
When charge is switched off ($q=0$), then Eqs.~(\ref{eta}) and (\ref{sumxi}) reduce to
\begin{eqnarray}
\eta &=& \frac{(x+a^2) \left[2 a^4+x (-5+6 x)+a^2 (1+8 x)\right] +2[x-(x+a^2)^2] 
\sqrt{2(x+a^2)}}
{\left(-1+2 a^2+2 x\right)^2},
\end{eqnarray}
\begin{eqnarray}
\xi_1 +\xi_2 &=& \frac{\left[a^2 -(x+a^2)^2\right]+\left[x-(x+a^2)^2\right]
\sqrt{2(x+a^2}}{a\left(-1+2 a^2+2 x\right)},
\end{eqnarray}
which are same as obtained for the five-dimensional Myers-Perry black holes 
\cite{Papnoi:2014}.

\section{five-dimensional EMCS Black hole shadow}
\label{5DShd}
Higher dimensions admit astrophysical objects such as supermassive black holes which are 
rather different from standard ones. The gravitational lensing features for alternative 
gravity theories with extra dimensions are different from general relativity \cite{Zakharov}.  
Several tests were proposed to discover signatures of extra dimensions in supermassive black 
holes since the gravitational field may be different from the standard one in  general 
relativity. In particular, it was shown how measurements of the shadow can have constraints 
on parameters of higher-dimensional black hole\cite{Zakharov}. When a black hole is situated 
between an observer and bright object, the light reaches the observer after being deflected by 
the black hole's gravitational field; but some part of the photons emitted by the bright 
object ends up with falling into the black hole, and this means photons never reach the 
observer.   

Our aim is to calculate the boundary curve of the shadow and the existence of a photon surface 
around the five-dimensional EMCS black holes which is necessary step to obtain the shadow. The 
incoming photons toward the black hole may follow three possible trajectories; either they 
fall into the black hole or they scattered away from the black hole and the third possibility 
concerns the critical geodesics that are the circular orbits around the black hole at critical 
radius. These are known as unstable orbits of constant radius (located at $r=3M$ for the 
Schwarzschild black hole). These are responsible for the apparent shape of the shadow of the 
black hole. To visualize the apparent shape of black hole we use celestial coordinates 
$\alpha$ and $\beta$, which can be calculated by defining the orthonormal basis vectors 
\cite{Johannsen:2015qca} for the local observer,
\begin{eqnarray}\label{loc_bas}
e_{\hat{t}} &=& \lambda e_t + \varsigma e_{\phi} + \chi e_{\psi}, \nonumber\\
e_{\hat{r}} &=& \frac{1}{\sqrt{g_{rr}}} e_r, \quad
e_{\hat{\theta}} = \frac{1}{\sqrt{g_{\theta \theta}}} e_{\theta}, \nonumber\\
e_{\hat{\phi}} &=& \frac{1}{\sqrt{g_{\phi \phi}}} e_{\phi}, \quad
e_{\hat{\psi}} = \frac{1}{\sqrt{g_{\psi \psi}}} e_{\psi},
\end{eqnarray}
where $\lambda$, $\varsigma$ and $\chi$ are constants and these are chosen in such a manner 
that the local basis vectors are orthogonal. The coefficients in Eq.~(\ref{loc_bas}) are real 
and one can verify that $\lbrace e_{t}, e_{r}, e_{\theta}, e_{\phi}, e_{\psi} \rbrace$ are 
orthonormal \cite{Johannsen:2015qca}. The basis vectors of the  local observer and the basis 
vectors of the metric (\ref{metric}) are related by
\begin{eqnarray}\label{relation}
e_{\hat{i}} &=&  e^{\mu}_{\hat{i}} e_{\mu}, \quad \hbox{and} 
\quad e^{\mu}_{\hat{i}} e^{\nu}_{\hat{j}} g_{\mu \nu} = \eta_{\hat{i} \hat{j}},
\end{eqnarray}
where $\eta_{\hat{i} \hat{j}}=(-1,1,1,1,1)$. With the help of Eqs.~(\ref{loc_bas}) and 
(\ref{relation}) and using the orthonormality condition of the basis vectors, one can obtain 
the constants $\lambda, \varsigma, \chi$ in the following form:
\begin{eqnarray}
\lambda &=& \frac{\sqrt{g_{\phi \psi}^2-g_{\phi \phi} g_{\psi \psi}}}
{\sqrt{g_{tt} g_{\phi \phi} g_{\psi \psi}+2 {g_{t\phi} g_{t\psi} g_{\phi \psi}-g_{t\phi}^2 
g_{\psi \psi}-g_{t\psi}^2 g_{\phi \phi} -g_{\phi \psi}^2 g_{tt}}}}, \nonumber\\
\varsigma &=& \frac{g_{t\phi} g_{\psi \psi}-g_{t\psi} g_{\phi \psi}}
{\sqrt{g_{tt} g_{\phi \phi} g_{\psi \psi}+2 {g_{t\phi} g_{t\psi} g_{\phi \psi}-g_{t\phi}^2 
g_{\psi \psi}-g_{t\psi}^2 g_{\phi \phi} -g_{\phi \psi}^2 g_{tt}}}}, \nonumber\\
\chi &=& \frac{g_{t\phi} g_{\phi \psi}-g_{t\psi} g_{\phi \phi}}
{\sqrt{g_{tt} g_{\phi \phi} g_{\psi \psi}+2 {g_{t\phi} g_{t\psi} g_{\phi \psi}-g_{t\phi}^2 
g_{\psi \psi}-g_{t\psi}^2 g_{\phi \phi} -g_{\phi \psi}^2 g_{tt}}}},
\end{eqnarray}
where the metric components are defined in Eq.~(\ref{components}). Further, the contravariant 
components of the three-momenta in the new coordinate basis \cite{Johannsen:2015qca} are 
given by 
\begin{eqnarray}\label{thmom}
p^{\hat{t}} &=& \lambda E - \varsigma L_{1} - \chi L_{2}, \nonumber\\
p^{\hat{\phi}} &=&  \frac{1}{\sqrt{g_{\phi \phi}}} L_{1}, \nonumber\\
p^{\hat{\psi}} &=& \frac{1}{\sqrt{g_{\psi \psi}}} L_{2}, \nonumber\\
p^{\hat{\theta}} &=& \frac{p_{\theta}}{\sqrt{g_{\theta \theta}}} 
= \frac{\pm\sqrt{\Theta}}{\sqrt{g_{\theta \theta}}}.
\end{eqnarray}
To describe the black hole shadow, we introduce the celestial coordinates 
\cite{Johannsen:2015qca}, which in five-dimensional black hole case takes the following form:
\begin{eqnarray}\label{def_cel}
\alpha &=& \lim_{r_{0}\rightarrow \infty} -r_{0} \frac{(p^{\hat{\phi}}+p^{\hat{\psi}})}
{p^{\hat{t}}},
\nonumber\\
\beta &=& \lim_{r_{0}\rightarrow \infty} r_{0} \frac{p^{\hat{\theta}}}{p^{\hat{t}}}.
\end{eqnarray}
Here $r_0$ is the distance from the black hole to observer, the coordinate $\alpha$ is the 
apparent perpendicular distance between the image and the axis of symmetry and the coordinate 
$\beta$ is the apparent perpendicular distance between the image and its projection on the 
equatorial plane \cite{Papnoi:2014}. We take the limit $r_0\rightarrow\infty$, since the 
observer is far away from the black hole. Also $\theta_0$ is the angular coordinate of the 
observer or the inclination angle. Substituting the contravariant components of three-momenta 
from Eq.~(\ref{thmom}) into  Eq.~(\ref{def_cel}), the celestial coordinates take the form
\begin{eqnarray}\label{celes}
\alpha &=& -\left( \xi_{1} \csc \theta_0 + \xi_{2}\sec \theta_0 \right), \nonumber\\
\beta &=& \pm \sqrt{\eta - \xi_{1}^2 \csc^2 \theta_0 - \xi_{2}^2 \sec^2 \theta_0 + a^2}.
\end{eqnarray}
Interestingly, Eq.~(\ref{celes}) has the same mathematical form as the five-dimensional 
Myers-Perry black hole \cite{Papnoi:2014} with modified impact parameters $\eta$ and 
$\xi_1 + \xi_2$  given respectively, by Eqs.~(\ref{eta}) and (\ref{sumxi}). However, the 
Eq.~(\ref{celes}) is different  from the Kerr-Newman black hole \cite{Takahashi:2005hy} with 
additional terms due to the extra dimension. Now we consider the case when an observer is 
situated in the equatorial plane of five-dimensional EMCS black hole, i.e., the inclination 
angle is $\theta_0=\pi/2$. In this case, the impact parameter $L_{\psi}=0$, therefore 
$\xi_2 =0$; hence  Eq.~(\ref{celes}) transforms to
\begin{eqnarray}\label{thpi2}
\alpha &=& -\xi_1, \nonumber\\
\beta &=& \pm \sqrt{\eta -\xi^2_1 +a^2}.
\end{eqnarray}
Similarly for $\theta_{0}=0$. In this case $L_{\phi}=0$ and hence $\xi_1 =0$,
\begin{eqnarray}\label{th0}
\alpha &=& -\xi_2, \nonumber\\ 
\beta &=& \pm \sqrt{\eta -\xi^2_2 +a^2}.
\end{eqnarray}
The shadows of a five-dimensional EMCS black hole can be visualized by plotting $\alpha$ vs. 
$\beta$ for different values of the rotation parameter ($a$) and the charge ($q$) at different 
inclination angles. The celestial coordinates $\alpha$ and $\beta$ in Eqs.~(\ref{thpi2}) and 
(\ref{th0}) satisfy the relation $\alpha^2 + \beta^2 = \eta + a^2$, where $\eta$ is given by 
Eq.~(\ref{eta}). It is clear that $\alpha$ and $\beta$ depend on charge $q$ and spin $a$.
\begin{figure}
	 \includegraphics[width=0.45\linewidth]{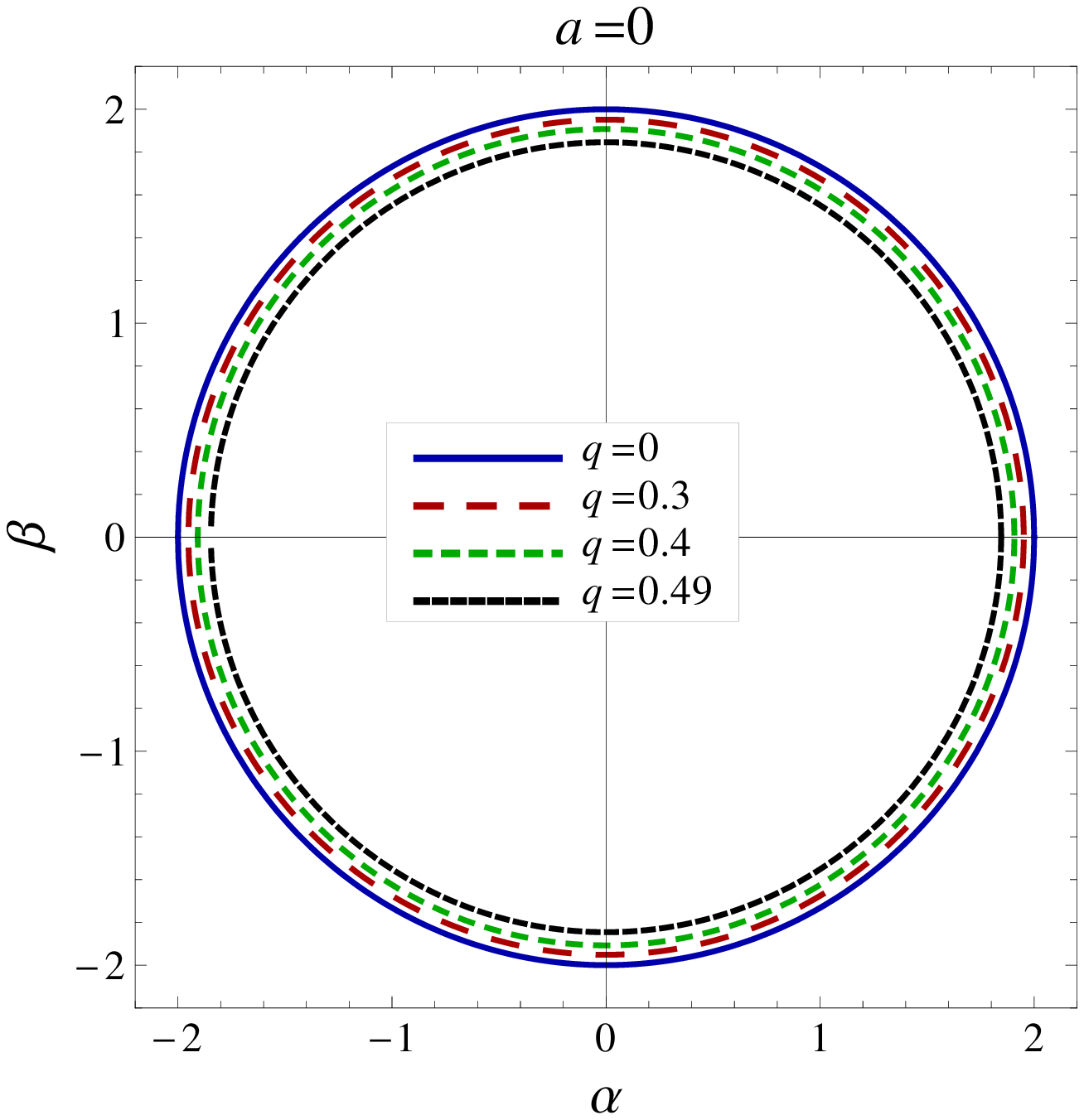}
	 \includegraphics[width=0.5\linewidth]{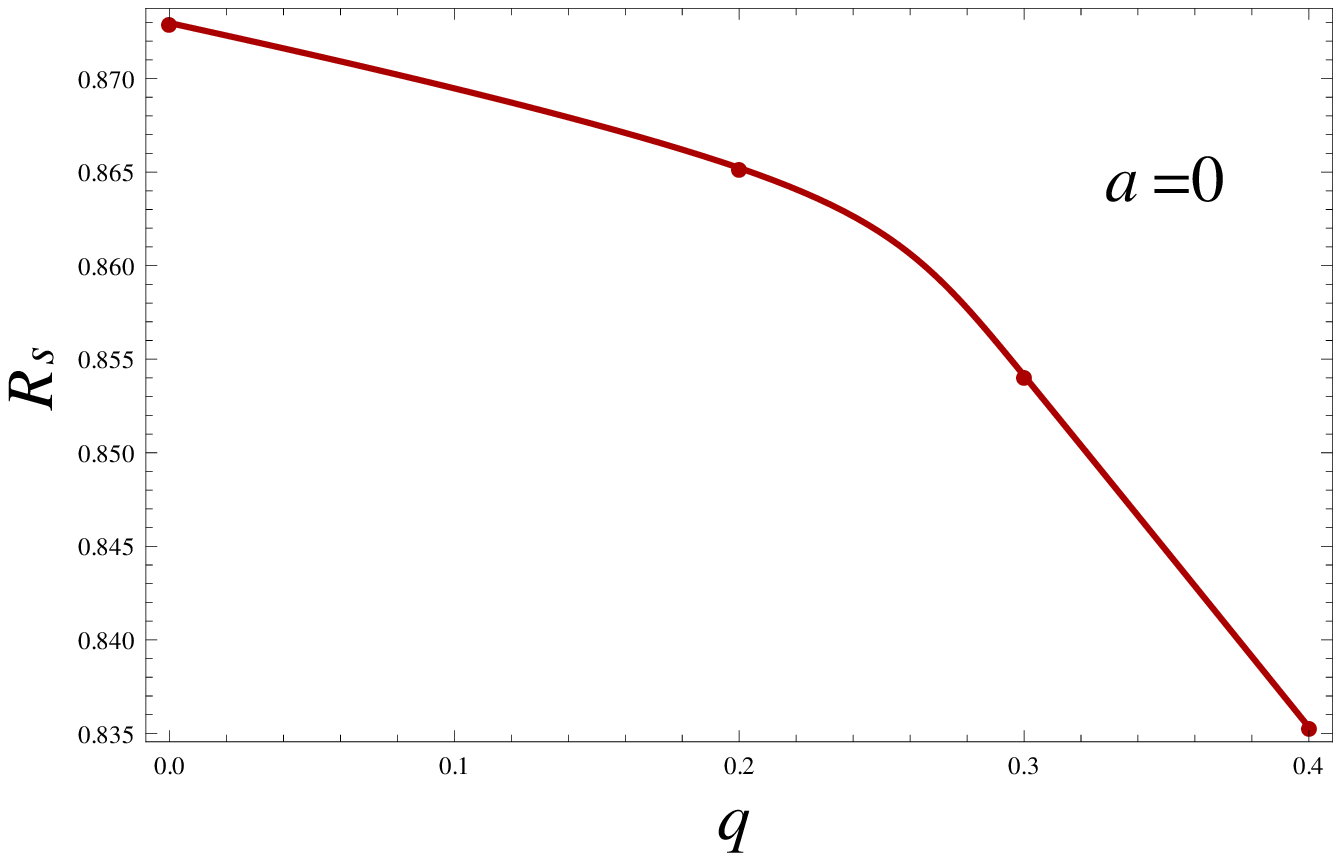}
	 \caption{\label{aeq0} Plot showing the shapes of black hole shadow cast by EMCS black 
	 hole with inclination angle $\theta_0=\pi/2$ and corresponding observable $R_s$, for 
	 different values of charge $q$ for the nonrotating case ($a=0$)}
\end{figure} 

The shadow with  $a=0$ for a five-dimensional nonrotating black hole can be obtained from
\begin{equation}\label{nonr}
\alpha^2 + \beta^2 = \frac{2 -9q^2 +2 (1-3q^2)^{3/2}}{1-4q^2} \equiv r^2_s \quad \text{with}  
\quad x = 1+\sqrt{(1-3q^2)}.
\end{equation}
The nonrotating five-dimensional EMCS black hole is a general case of five-dimensional 
Reissner-Nordstr{\"o}m black hole and its shadow appears as a perfect circle with radius 
$R_s$ (cf. Fig.~\ref{aeq0}). We plotted the shadow of a nonrotating five-dimensional EMCS 
black hole for several  values of charge $q$. The effect of charge $q$ can be seen by 
the radius of the circle with an increase in $q$ (cf. Fig.~\ref{aeq0}). When $q=0$, the 
radius of the shadow  is $R_s=2$, which is similar to the five-dimensional Schwarzschild black 
hole \cite{Singh:2017vfr}. Thus, the effect of charge $q$ is to decrease the size of the 
shadow (cf. Fig.~\ref{aeq0}).
\begin{figure*}
	\begin{tabular}{c c c c}
	 \includegraphics[width=0.5\linewidth]{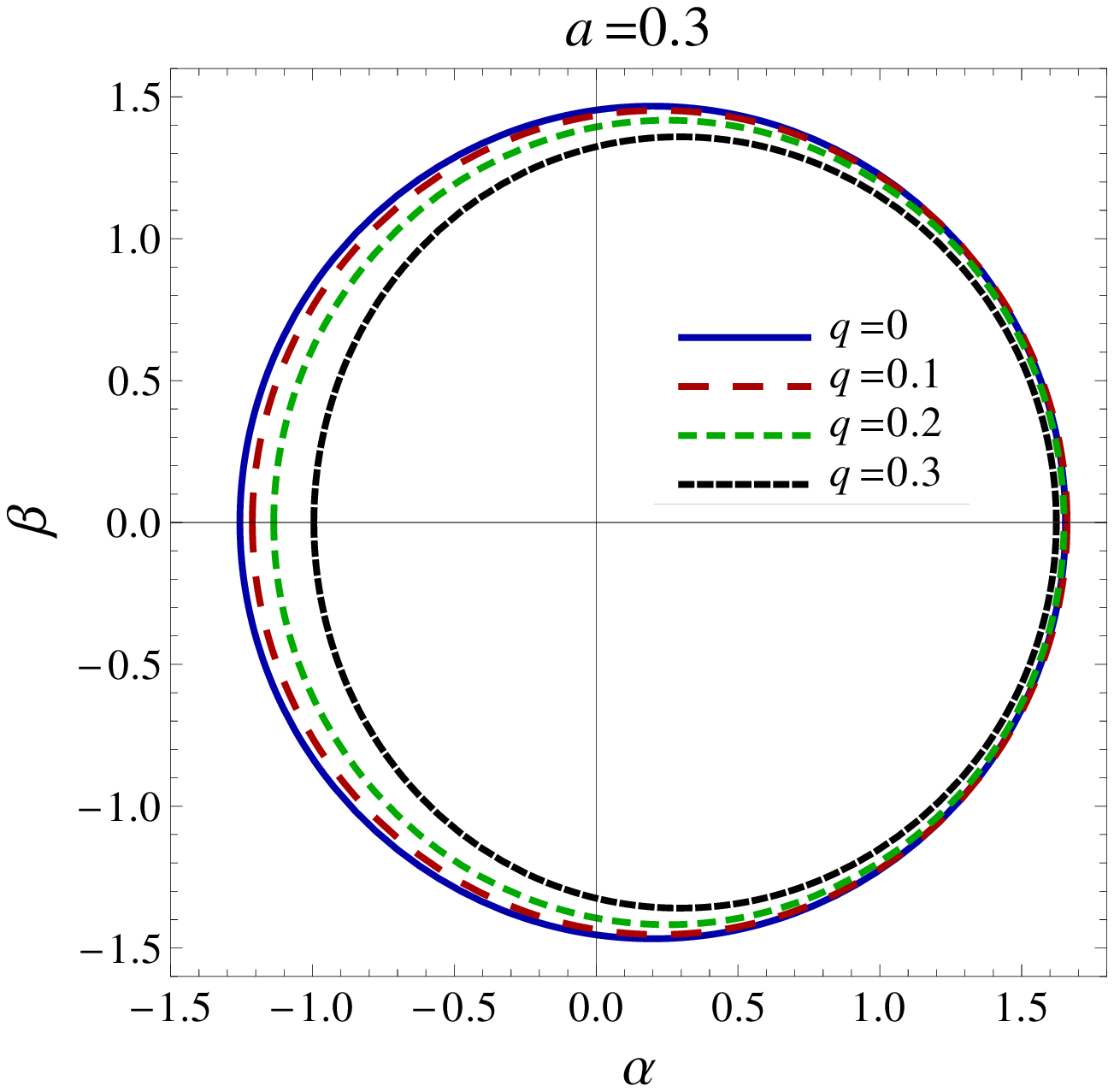}
	 \includegraphics[width=0.48\linewidth]{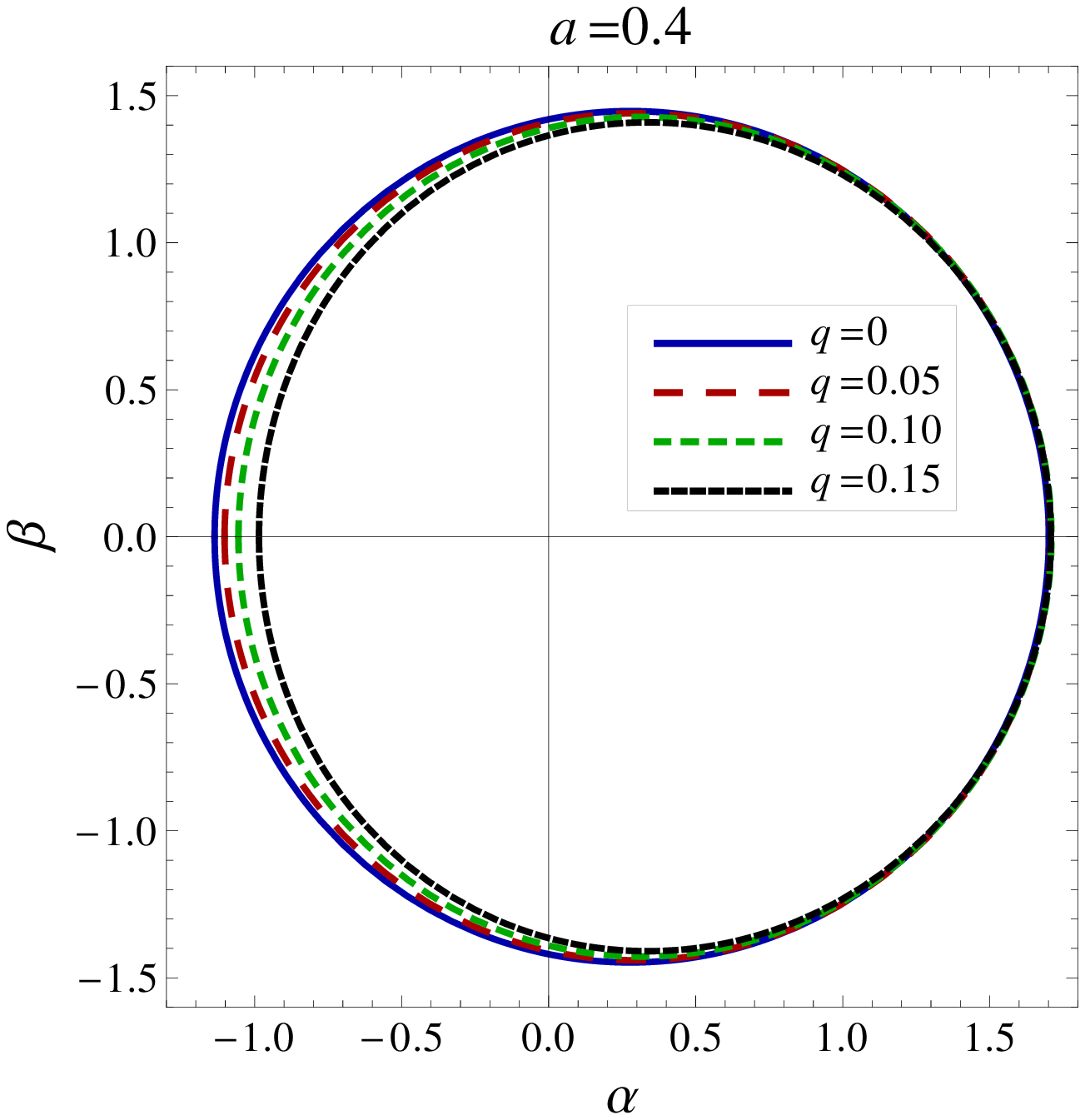}\\
	 \includegraphics[width=0.5\linewidth]{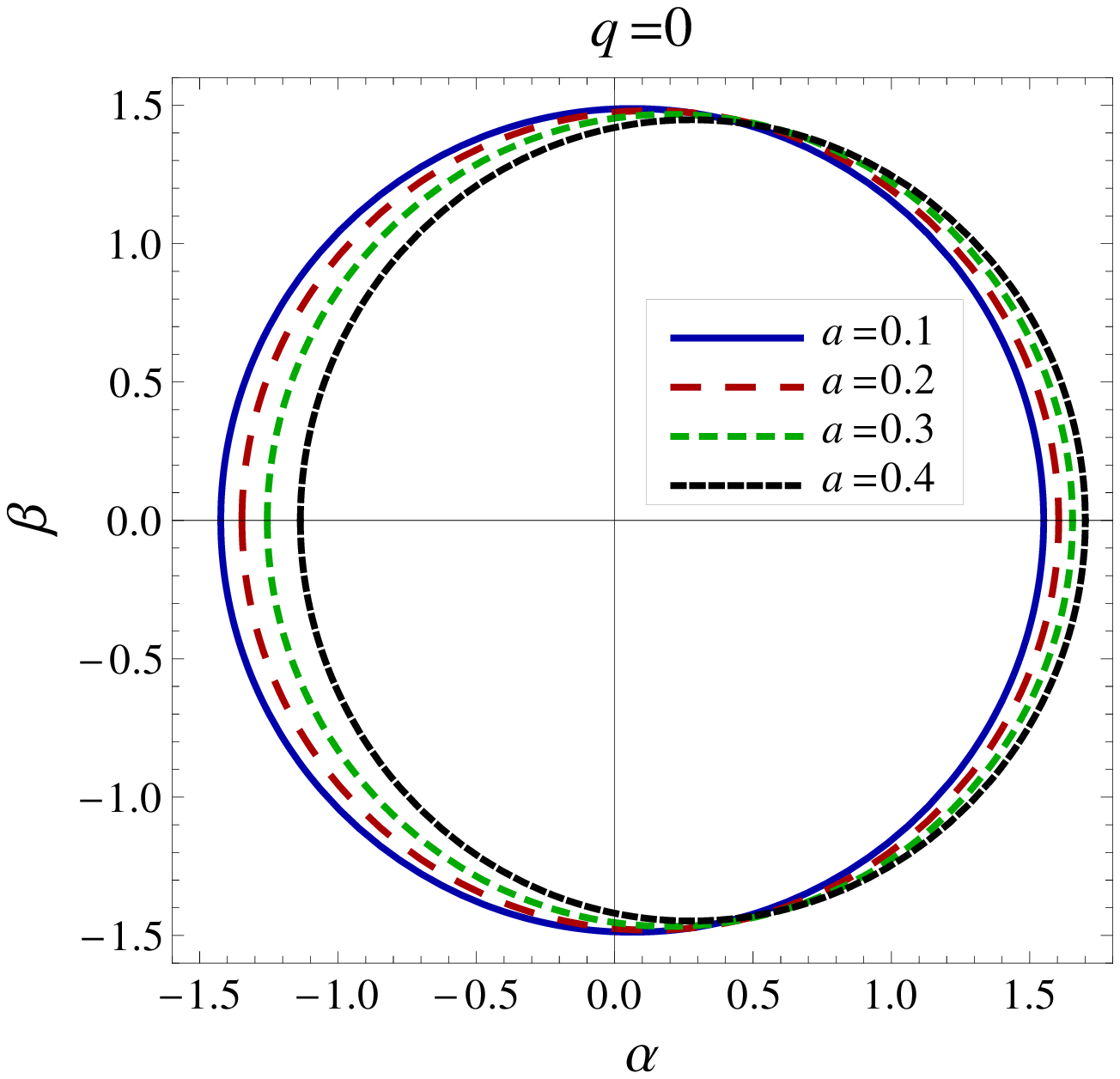}
	 \includegraphics[width=0.5\linewidth]{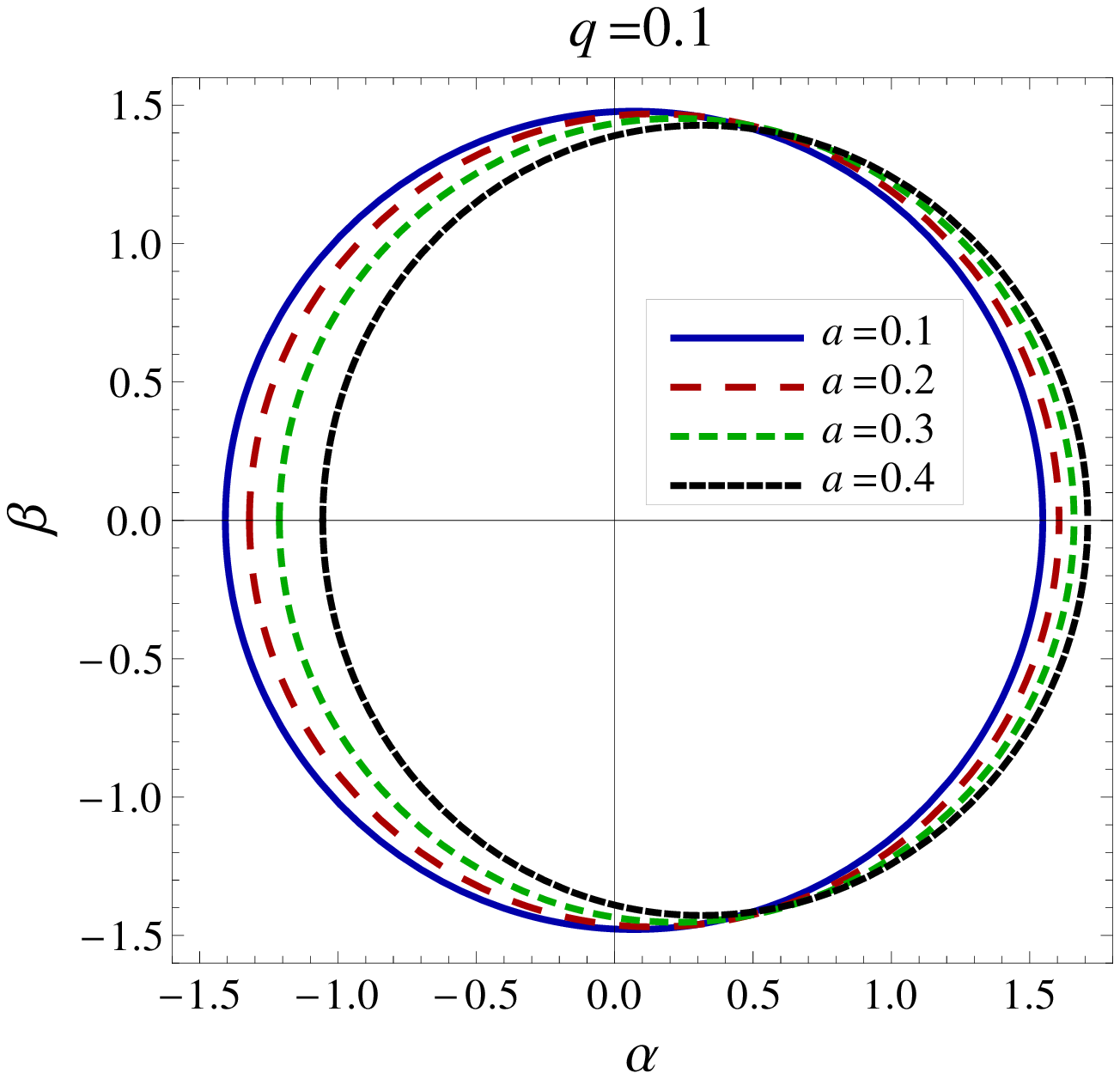}
	\end{tabular}
\caption{\label{aneq0} Plot showing the shapes of shadows cast by five-dimensional EMCS black 
hole with inclination angle $\theta_0=\pi/2$ for different values of charge $q$ and the spin 
$a$ with the case $q=0$ referring to Myers-Perry black hole}
\end{figure*} 

Next, we consider the rotating case of the five-dimensional EMCS black hole to see the 
behavior of black hole shadow in the presence of both spin $a$, charge $q$, and extra 
dimension. With increasing $a$, the shadow gets more and more distorted and shifts to 
rightmost on the vertical axis, as in the Kerr case \cite{Hioki:2009na}. In the absence 
of charge $q=0$, one gets 
\begin{equation}\label{alpha+beta}
\alpha^2 + \beta^2 = \frac{(x+ a^2) \left[2 a^4+x (-5+6 x)+a^2 (1+8 x)\right] 
+2 \left(x-(x+a^2)^2\right) \sqrt{2(x+a^2)}}{(-1+2 a^2+2 x)^2}+a^2.
\end{equation}
The shapes of shadow for the five-dimensional EMCS black hole have been plotted in 
Fig.~\ref{aneq0} for different values of charge $q$ and spin $a$. The shape of the black hole 
shadow is a deformed circle instead of a perfect circle. We see that the shape of the shadow 
is largely affected due the parameters $a$, $q$ and extra dimension. The size of shadow 
decreases continuously (cf. Fig.~\ref{aneq0}) with  increase in $q$. This can be understood 
as a dragging effect due to the rotation of the black hole. The extra dimension  has same 
effect on the size of the shadow\cite{Papnoi:2014}. 
\begin{figure}
	\includegraphics[scale=0.42]{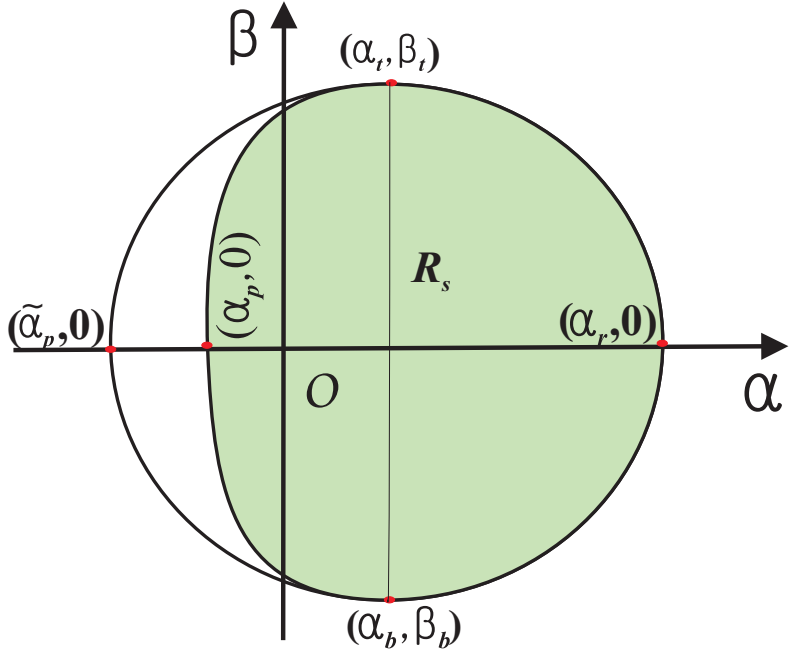}
    \caption{\label{observable} Schematic representation of the observables for rotating 
    black holes \cite{Amir:2016cen}}
\end{figure}

Next, it will be helpful to introduce observables which characterize the shape and the 
distortion of the shadow.  The characterization of the observables useful to extract more 
information from the black hole shadow \cite{Schee:2008kz}. We modify the Hioki-Maeda 
characterization \cite{Hioki:2009na} to determine the observables, radius $R_s$ and distortion 
$\delta_s$ for rotating five-dimensional EMCS black hole \cite{Hioki:2009na,Amarilla:2011fxx}:
\begin{eqnarray}
R_s &=& \frac{(\alpha_t - \alpha_r)^2 + \beta_t^2}{2(\alpha_t - \alpha_r)}, 
\end{eqnarray}
and
\begin{eqnarray}
\delta_{s} &=& \frac{\tilde{\alpha_p} -\alpha_p}{R_s},
\end{eqnarray}
where $(\tilde{\alpha_p} ,0)$ and $(\alpha_p,0)$ are the coordinates where the reference 
circle and the contour of the shadow cut the horizontal axis on the opposite side of 
$(\alpha_r,0)$ (cf. Fig.~\ref{observable}). In this characterization, the idea was that a 
reference circle is passing through the three coordinates of the black hole shadow. The 
coordinates are situated at top position ($\alpha_t$, $\beta_t$), bottom position 
($\alpha_b$, $\beta_b$), and rightmost position ($\alpha_r$, 0) (cf. Fig.~\ref{observable}). 
The behavior of the radius $R_s$ and the distortion $\delta_s$ with charge for different 
values of spin $a$ is depicted in Fig.~\ref{obs}, which suggests that the radius $R_s$ 
monotonically decreases as $q$ increases, and the distortion $\delta_s$ of the shadow 
increases with $q$ and the shadow gets more distorted for larger values of $a$. When compared 
with the five-dimensional Myers-Perry black hole \cite{Papnoi:2014}, the effective size of the 
shadow decreases for higher values of the charge $q$. Also, a comparison with the shadow with 
Kerr-Newman black holes \cite{Takahashi:2005hy} indicates that the effective size of the 
shadow decreases due to the extra dimensions. Instead of using $R_s$ and $\delta_s$, one can 
also use the observables defined by Schee and Stuchl{\'i}k \cite{Schee:2008kz} to arrive at 
the same conclusions.
\begin{figure*}
    \begin{tabular}{c c c c}
	\includegraphics[scale=0.6]{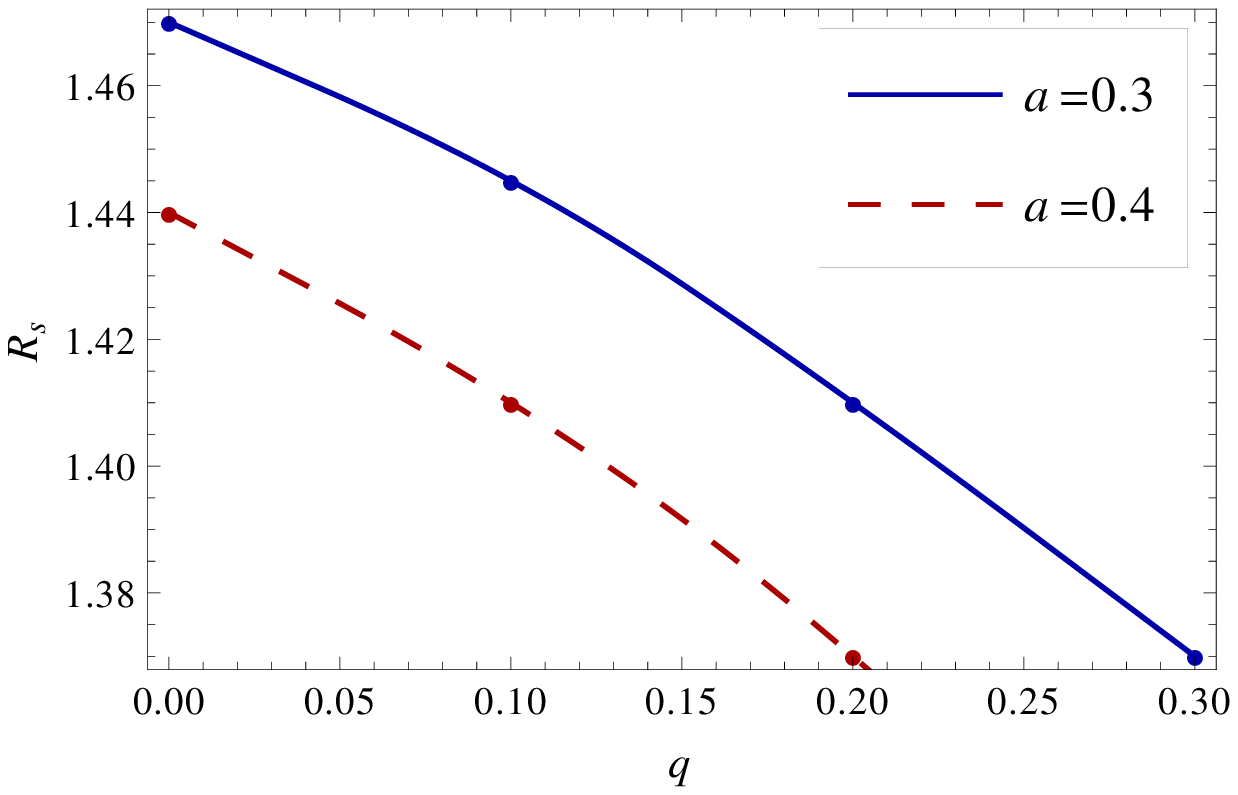} 
	\includegraphics[scale=0.6]{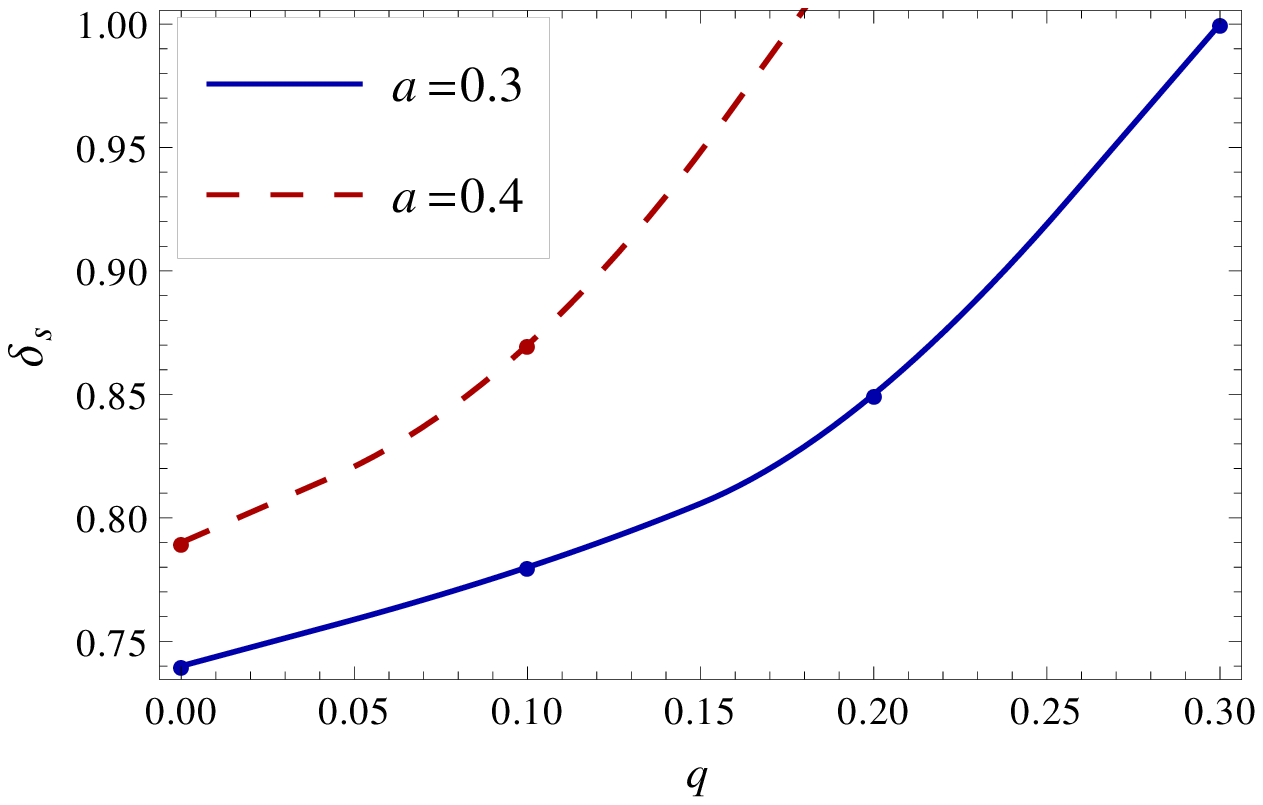} \\
   	\end{tabular}
    \caption{\label{obs} Plots showing the variation of radius of black hole shadow $R_s$ and 
    distortion parameter $\delta_s$ with charge $q$ for different value of spin $a$ }
\end{figure*}
\begin{figure*}
	\begin{tabular}{c c c c}
	 \includegraphics[width=0.5\linewidth]{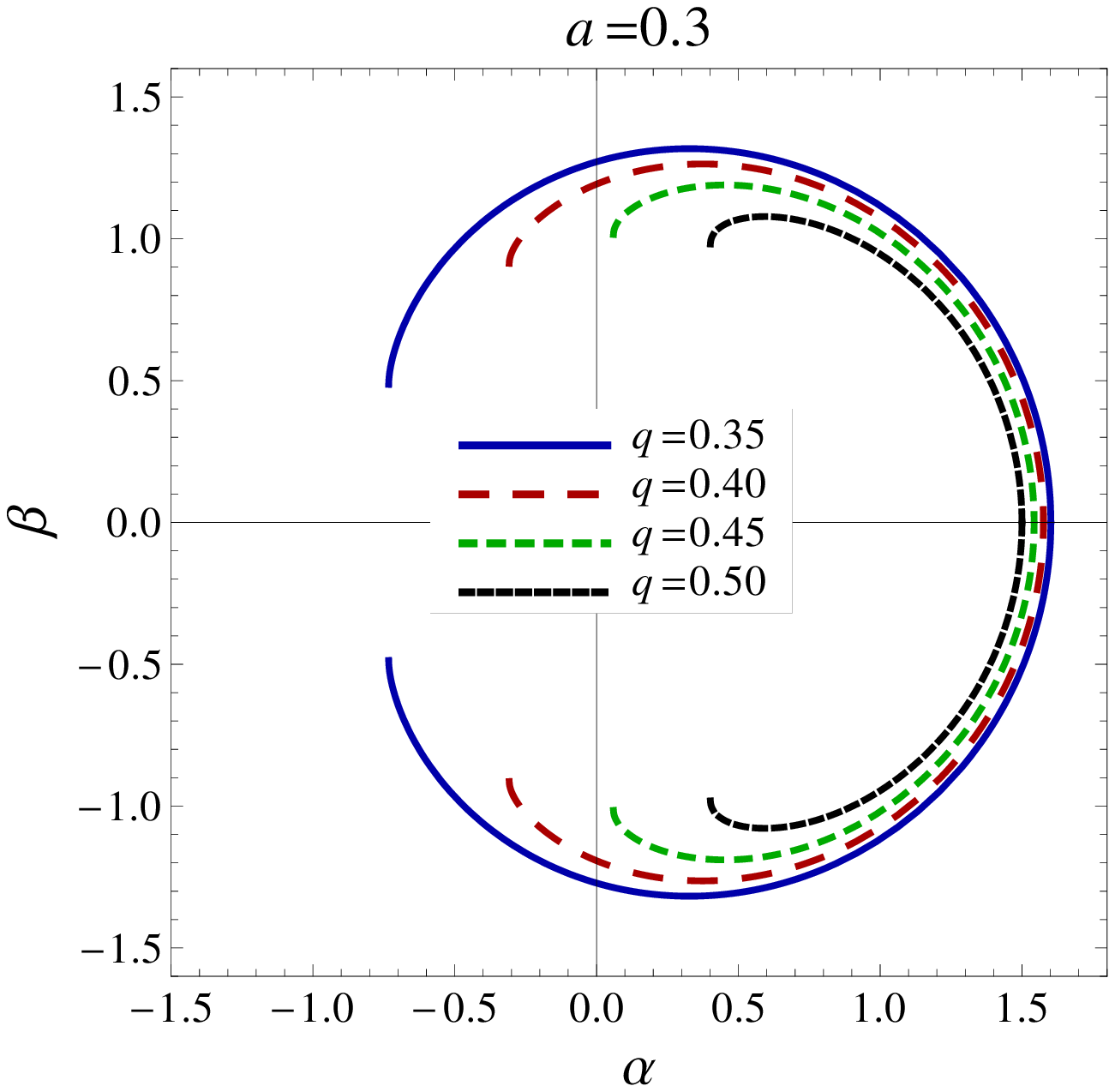}
	 \includegraphics[width=0.48\linewidth]{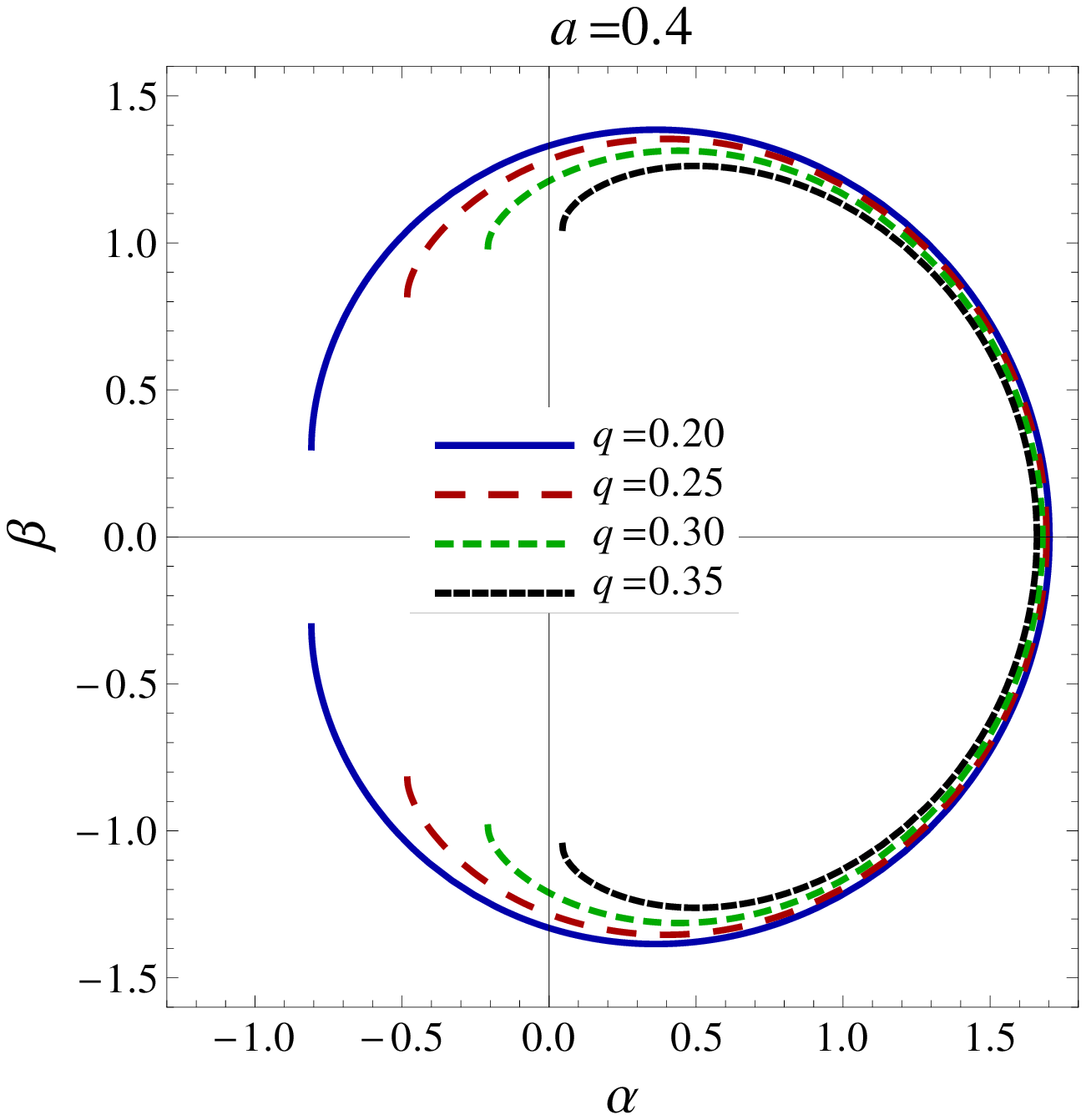}\\
	 \includegraphics[width=0.5\linewidth]{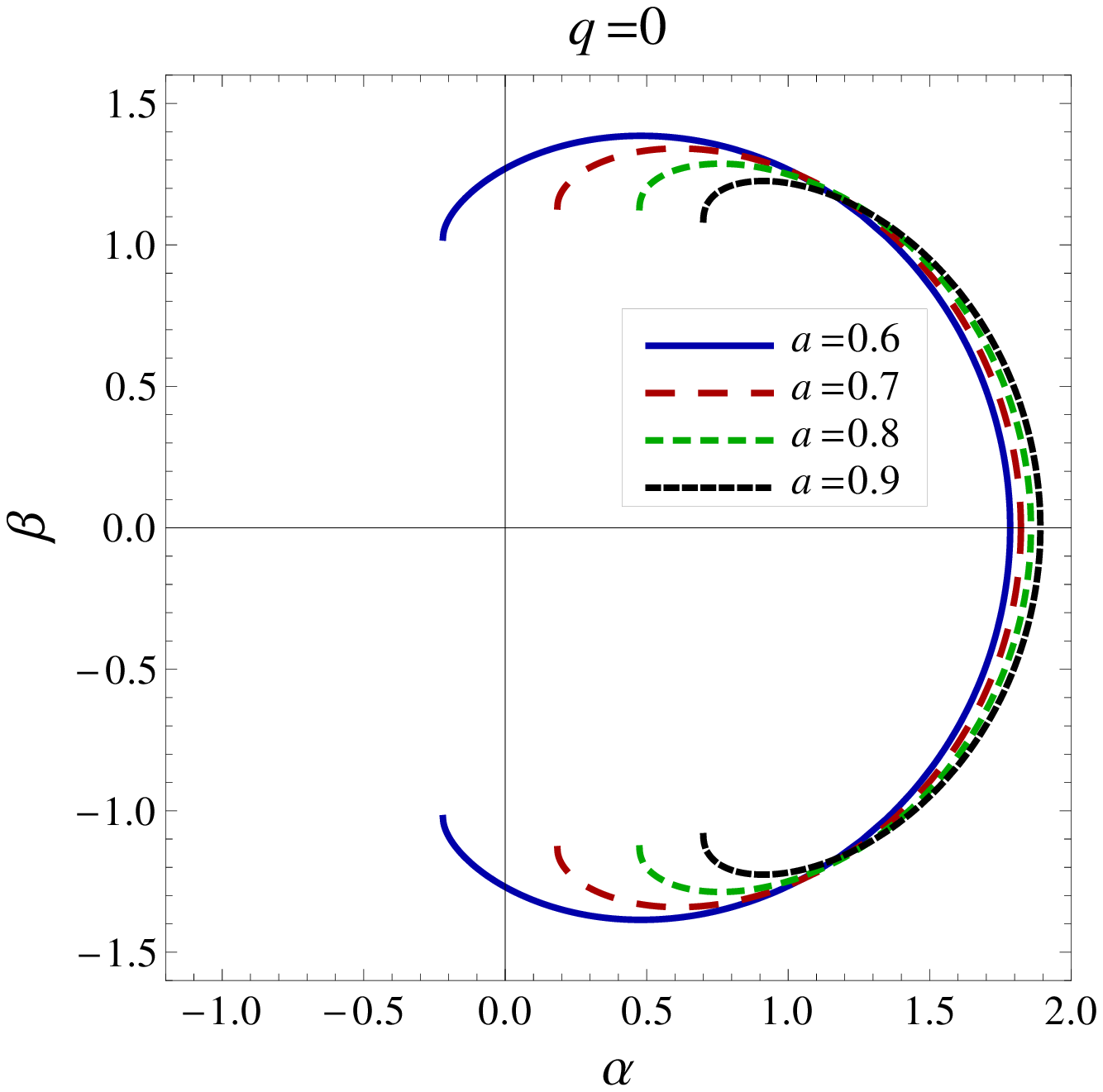}
	 \includegraphics[width=0.5\linewidth]{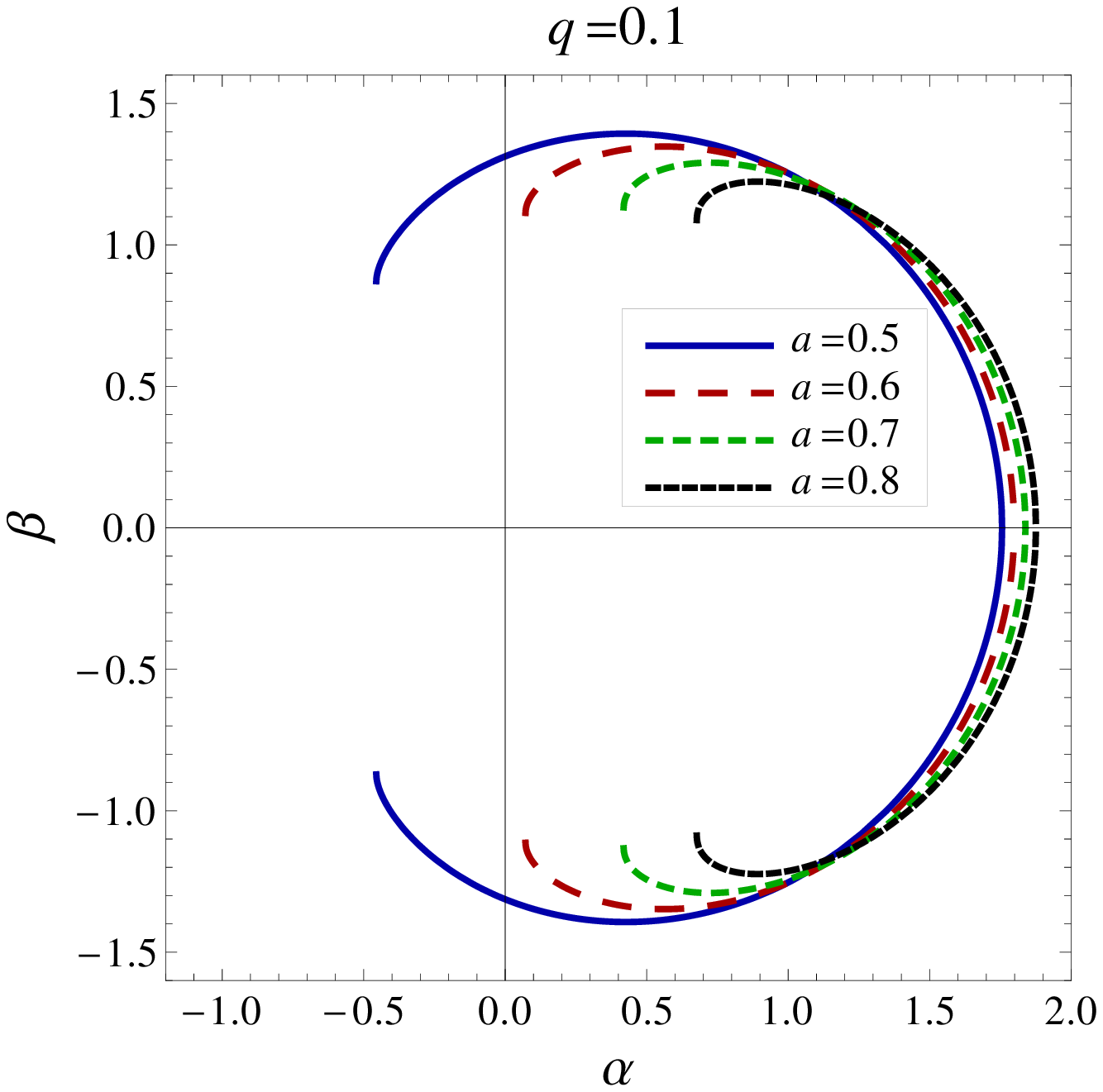}
	\end{tabular}
\caption{\label{naked} Shadow cast by the naked singularity with inclination angle 
$\theta_0=\pi/2$ for different values of the charge $q$ and the spin $a$. Here Myers-Perry 
black hole ($q=0$) case is included for comparison}
\end{figure*}

\section{Naked singularity}
\label{naksing}
The cosmic censorship hypothesis states that the spacetime singularities are always hidden by 
the event horizon of the black hole \cite{Penrose:1969pc}. However, a naked singularity can be 
defined as a gravitational singularity in the absence of the event horizon, which leads to a 
violation of cosmic censorship hypothesis \cite{Penrose:1969pc}. The cosmic censorship 
hypothesis has as yet no precise mathematical formulation or proof for either version and 
remains one of the most important unsolved problems in  general relativity. Consequently, the 
examples that appear to violate the cosmic censorship hypothesis are important and these are 
useful tools for the study of this crucial issue. Recently, Figueras {\it et al.} 
\cite{Figueras:2015hkb} have found the strongest evidence for a violation of the weak cosmic 
censorship conjecture in five-dimensional spacetime. Hence, it is important to study the naked 
singularity shadow for the five-dimensional EMCS black hole.

A five-dimensional EMCS black hole admits a naked singularity, it occurs when $\mu<(a+b)^2+2q$ 
or for the case of $a=b$, $\mu<2(2a^2+q)$. In the absence of charge $q$, the condition for the 
existence of a naked singularity satisfy $\mu<(a+b)^2$ ($a \neq b$) and $\mu<4a^2$ ($a=b$). 
The  most general condition for naked singularity in the five-dimensional Myers-Perry 
spacetime is $a>1/{\sqrt{2}}$ \cite{Papnoi:2014}. The shadow of naked singularity for the 
five-dimensional Myers-Perry spacetime is studied in \cite{Papnoi:2014}. Here we plot the 
contour of the naked singularity shadows for different values of $q$ and $a$ in 
Fig.~\ref{naked}, and the shapes change dramatically   as compared to   black hole shadows. 
It is found that the shape of a naked singularity shadow forms an arc instead of a circle 
(cf. Fig.~\ref{naked}). The behavior is similar to the naked singularity in the Myers-Perry 
geometries \cite{Papnoi:2014} or in the braneworld case \cite{Amarilla:2011fxx}. The unstable 
spherical photon orbits with a positive radius give way to an open arc instead of a closed 
curve due to the absence of the event horizon and photons can reach the observer. The shapes 
of the naked singularity shadows are affected by charge $q$ as well as spin $a$; its size 
decreases in both  cases when we increase either $q$ or $a$ (cf. Fig.~\ref{naked}). Our 
results show that the naked singularity shadow is different from the five-dimensional 
Myers-Perry black hole \cite{Papnoi:2014} (cf. Fig.~\ref{naked}). We observe that the arc of 
the shadow tends to open with the increasing values of the spin $a$ as well as charge $q$. 
The two observables, viz. $R_s$ and $\delta_{s}$  are no longer valid for a naked singularity 
and new observables are necessary.

\section{Energy emission rate}
\label{emission}
Here, we study the energy emission rate of the five-dimensional EMCS black holes, as in the 
Kerr black holes \cite{Atamurotov:2015nra}. It is well known that the shadow is responsible 
for a high energy absorption cross section due to the black hole for a far away observer 
\cite{Misner:1973}. The energy emission rate of a black hole can be calculated using the 
following relation \cite{Mashhoon:1973zz,Wei:2013kza}:
\begin{equation}
\frac{d^2E(\omega)}{d\omega dt}= \frac{2 \pi^2 \sigma_{lim}}{\exp{(\omega/T)}-1}\omega^3,
\end{equation}
where $\omega$ is the frequency, $\sigma_{lim}$ is the limiting constant value for a 
spherically symmetric black hole around which the absorption cross section oscillates, and 
$T$ is the Hawking temperature of the black hole. The Hawking temperature of the 
five-dimensional EMCS black hole reads \cite{Wu:2009cn}
\begin{equation}\label{temp}
T=\frac{{(x^{H}_{+})}^2-(a^2+q)^2}
{2 \pi \sqrt{x^{H}_{+}}\left[a^2 q +(x^{H}_{+}+a^2)^2\right]},
\end{equation}
where $x^{H}_{+}$ is the event horizon. Interestingly, the Hawking temperature (\ref{temp}) 
depends on the charge $q$ as well as the spin $a$. When $q \rightarrow 0$, the temperature 
reduces to
\begin{equation}
T = \frac{x^{H}_{+}-a^2}{2 \pi \sqrt{x^{H}_{+}} (x^{H}_{+}+a^2) },
\end{equation}
which shows the temperature of the Myers-Perry black holes \cite{Altamirano:2014tva}. The 
limiting constant value of a five-dimensional EMCS black hole can be expressed approximately  
\cite{Mashhoon:1973zz,Misner:1973} by
\begin{equation}
\sigma_{lim} \approx \pi R_s^3,
\end{equation}
where $R_s$ is the radius of the black hole shadow. Hence, the complete form of the energy 
emission rate for a five-dimensional EMCS black hole is
\begin{eqnarray}
\frac{d^2E(\omega)}{d\omega dt}=\frac{4\pi^3 R_s^3}{e^{\omega/T}-1}\omega^3.
\end{eqnarray}
\begin{figure*}
    \begin{tabular}{c c c c}
	\includegraphics[scale=0.5]{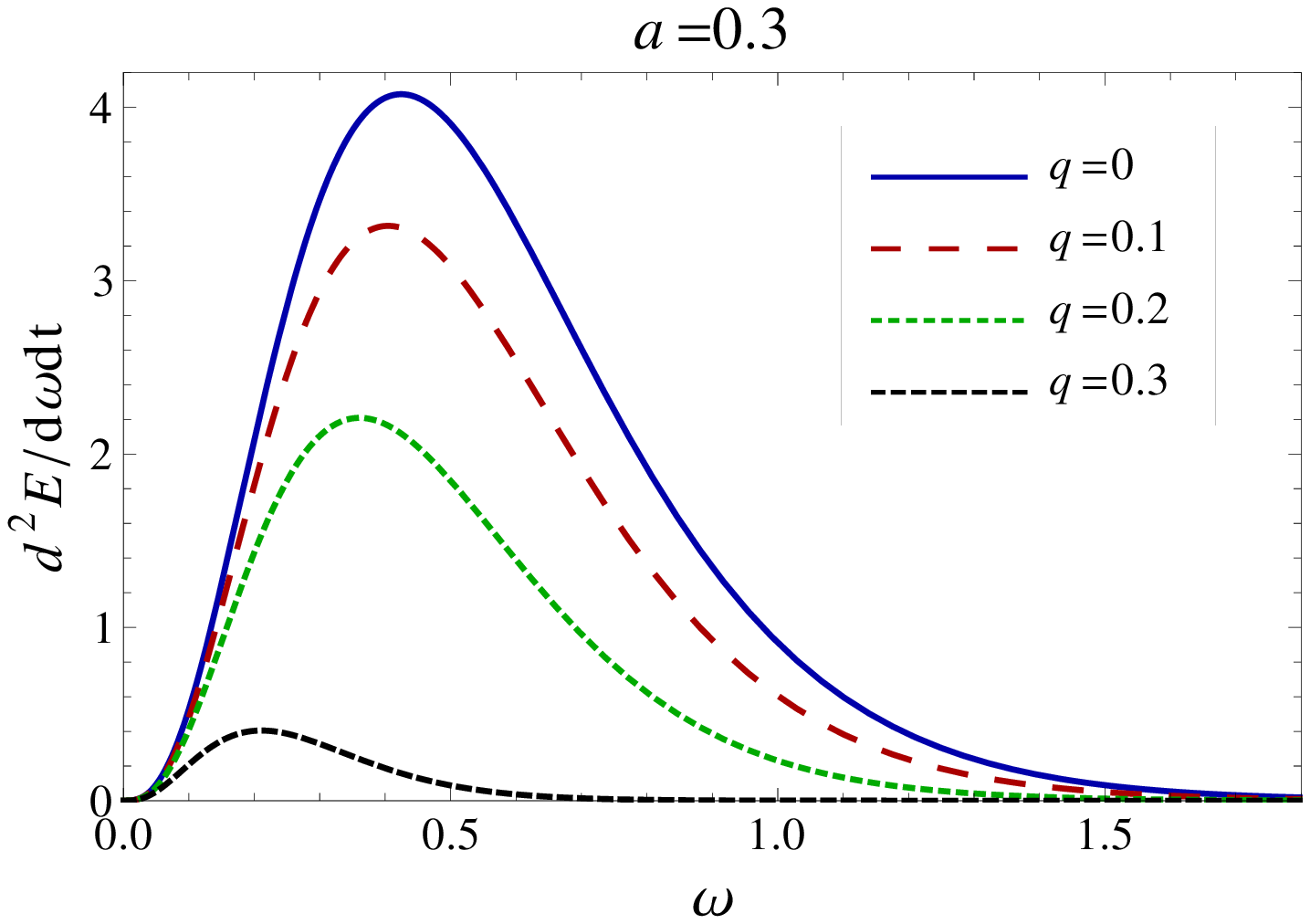} 
	\includegraphics[scale=0.5]{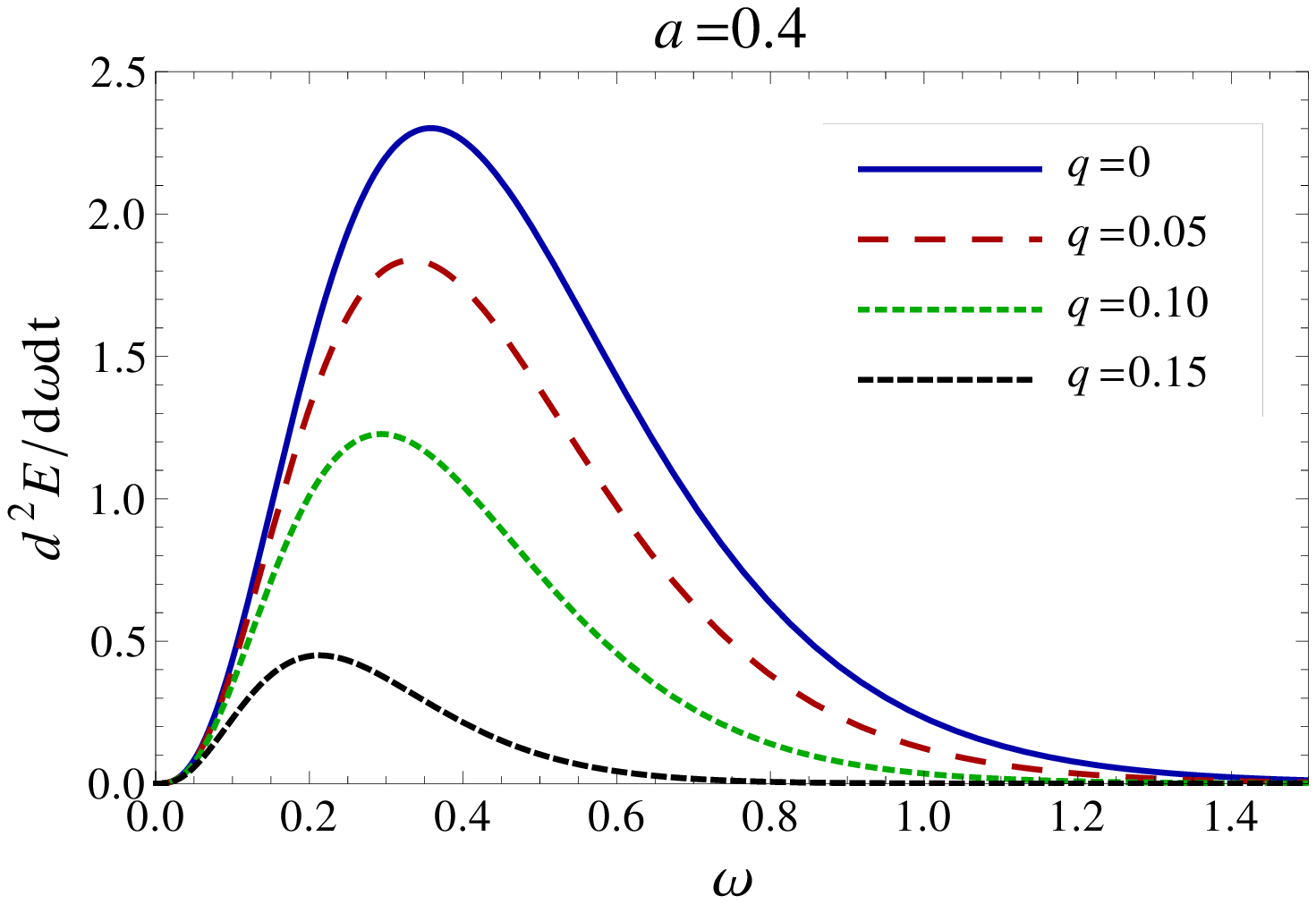} \\
   	\end{tabular}
    \caption{\label{emsnr} Plots showing the variation of the energy emission rate with 
    frequency $\omega$ for different values of charge $q$ and spin $a$ for the 
    five-dimensional EMCS black holes with case $q=0$ refers to Myers-Perry black hole}
\end{figure*}
In Fig.~\ref{emsnr}, we plot the energy emission rate versus frequency $\omega$ for different 
values of the charge $q$ and the spin $a$. It can be seen that an increase in the values of 
$q$ and $a$ decreases the peak of energy emission rate (cf. Fig.~\ref{emsnr}). We analyze how 
the energy emission rate of a five-dimensional EMCS black hole decreases in comparison with a 
five-dimensional Myers-Perry black hole \cite{Papnoi:2014}.

\section{Conclusion}
\label{conclusion}
The formation of a shadow due to the strong gravitational field near a black hole has received 
significant attention due to a possibility of observing the images of supermassive black hole 
Sgr $A^*$ situated at the center of our galaxy \cite{falcke}. The gravitational theories with 
extra dimensions admit black hole solutions which have different properties from the standard 
ones. Several tests were proposed to discover the signatures of extra dimensions in black 
holes as the gravitational field likely to be distinct from the one in general relativity, 
e.g., the gravitational lensing for gravities with extra dimensions has different 
characteristics from general relativity. It turns out that the measurements of shadow sizes 
for higher dimensional black holes can put constraints on the parameters of black holes 
\cite{Zakharov}, e.g., evaluating the size of a shadow it was shown that the probability to 
have a tidal charge ($-q$) for the black hole at the galactic center is ruled out by the 
observations \cite{doeleman}. Thus, it suggests that black holes with positive charge ($+q$) 
are consistent with observations, but a significant negative charge ($-q$) black holes is 
ruled out \cite{doeleman}, i.e., the Reissner-Nordstr{\"o}m black holes which have positive 
charge ($+q$) are consistent with the observation, but this is not true for the braneworld 
black holes, which have a negative tidal charge ($-q$). A five-dimensional EMCS black hole 
solution has an additional charge parameter $q$ when compared with the five-dimensional 
Myers-Perry black hole, and it provides deviation from the Myers-Perry black hole. The 
five-dimensional EMCS black hole has a richer configuration for the horizons and ergosphere. 
It is interesting to note that the ergosphere size is sensitive to the charge $q$ as well as 
rotation parameter $a$. 

We have extended the previous studies of black hole shadow and derived analytical formulas for 
the photon regions for a five-dimensional EMCS black hole. We also make quantitative analyses 
of the shape and size of the black hole shadow cast by a rotating five-dimensional EMCS black 
hole. We have analyzed how the shadow of black hole is changed due to the presence of charge 
$q$, and we explicitly show that the charge $q$ apparently affects the shape and the size of 
the shadow. In particular, it is observed that the shadow of a rotating five-dimensional EMCS 
black is a dark zone covered by a more deformed circle as compared to Myers-Perry black hole 
shadow. For a fixed value of spin $a$, compared to the rotating five-dimensional Myers-Perry 
black hole, the size of the shadow decreases with the charge $q$, whereas shadows become more 
distorted with an increase in charge $q$, and the distortion is maximal for an extremal black 
hole. In comparison with the four-dimensional Kerr-Newman black hole, we found that the shadow 
of the five-dimensional EMCS black hole decreases. The study of the shadow of naked 
singularity  of the five-dimensional EMCS metric suggests that the shape of the shadow 
decreases for higher values of the charge $q$ when compared with the five-dimensional 
Myers-Perry metric. Obviously, our results in the limit $q\rightarrow 0$ reduce exactly to 
five-dimensional Myers-Perry black hole.

We think that the results obtained here  are of interest in the sense that they do offer the 
opportunity to explore  properties associated with a shadow in higher dimensions, which may 
be crucial in our understanding whether the size of shadow suggest if there are signatures of 
a five-dimensional charged black hole. The possibility of a further generalization by adding a 
negative cosmological constant is an interesting problem for future research under active 
consideration.

\begin{acknowledgements}
S.G.G. would like to thank SERB-DST Research Project Grant No. SB/S2/HEP-008/2014 and DST 
INDO-SA bilateral project DST/INT/South Africa/P-06/2016. M.A. acknowledges the University 
Grant Commission, India, for financial support through the Maulana Azad National Fellowship 
For Minority Students scheme (Grant No. F1-17.1/2012-13/MANF-2012-13-MUS-RAJ-8679). 
\end{acknowledgements}

\end{document}